\documentclass{iopart}
\usepackage{epsfig}
\def\lsim{\hbox{ \raise.35ex\rlap{$<$}\lower.6ex\hbox{$\sim$}\ }}
\def\gsim{\hbox{ \raise.35ex\rlap{$>$}\lower.6ex\hbox{$\sim$}\ }}

\def\xrightarrow#1#2#3#4{\,\lower#1pt\hbox{$\stackrel{\stackrel{\displaystyle #2}%
{\hbox to #3cm{\rightarrowfill}}}{#4}$}\,}

\def\setC{{\bf C}}
\def\setR{{\bf R}}

\begin{document}
\title{Constraints on Supersymmetric Grand Unified Theories from Cosmology}
\author{Jonathan Rocher$^{1\dagger}$, Mairi Sakellariadou$^{2\star}$}
\address{$^1$
Institut d'Astrophysique de Paris, 
{${\cal G}\setR\varepsilon\setC{\cal O}$}, 
FRE 2435-CNRS, 98bis
boulevard Arago, 75014 Paris, France\\
$^2$
Division of Astrophysics, Astronomy
and Mechanics, Department of Physics, University of Athens,
Panepistimiopolis, GR-15784 Zografos (Athens), Hellas\\
and  Institut d'Astrophysique de Paris, 98bis boulevard Arago, 
75014 Paris, France}
\ead{$^\dagger$rocher@iap.fr,
$^\star$msakel@cc.uoa.gr, sakella@iap.fr}

\begin{abstract}
Within the context of SUSY GUTs, cosmic strings are generically 
formed at the end of hybrid inflation. However, the WMAP CMB measurements
strongly constrain  the
possible cosmic strings contribution to the angular power spectrum
of anisotropies. We investigate the parameter space of
SUSY hybrid (F- and D- term) inflation, to get the conditions under
which  theoretical predictions are in agreement with  data.  
The predictions
of F-term inflation are in agreement with  data, only if the
superpotential coupling $\kappa$ is small. In particular, for 
SUSY SO(10), the upper bound is $\kappa\lsim
7\times 10^{-7}$. This fine tuning problem can  be lifted if we employ 
the curvaton
mechanism, in which case $\kappa\lsim 8\times 10^{-3}$; higher
values are not allowed by the gravitino constraint. The constraint on
 $\kappa$ is equivalent to  a constraint on the
SSB mass scale $M$, namely $M \lsim 2\times 10^{15}$ GeV. \\
The study of D-term inflation shows that the inflaton field is of the
order of the Planck scale;  one  should therefore consider SUGRA. 
We find that the cosmic strings contribution to the CMB
anisotropies is not constant, but it is strongly
dependent on the gauge coupling  $g$ and on the superpotential
coupling $\lambda$. We obtain
 $g\lsim 2\times 10^{-2}$ and $\lambda \lsim
3\times 10^{-5}$. SUGRA corrections induce also a lower limit 
for $\lambda$. Equivalently, the Fayet-Iliopoulos term $\xi$
must satisfy  $\sqrt\xi \lsim 2\times 10^{15}$ GeV.
This constraint holds for all allowed values of $g$.
\end{abstract}

\pacs{cos, ssb, suy}

\section{Introduction}

High energy physics and cosmology are two complementary areas with a
rich and fruitful interface. They both enter the description of the
physical processes during the early stages of our Universe.  High
energy physics leads to the notion of topological defects, which imply
 a number of cosmological consequences.  Once we compare the
theoretical predictions of models motivated by high energy physics
against cosmological data, we induce constraints or, in other words,
we fix the free parameters of the models.  This is the philosophy of
our study.

Even though the particle physics Standard Model (SM) has been tested
to a very high precision, evidence of neutrino
masses~\cite{SK,SNO,kamland} proves that one should go beyond this
model. An extension of the SM gauge group can be accomplished in the
framework of Supersymmetry (SUSY). At present, SUSY is the only viable
theory for solving the gauge hierarchy problem. Moreover, in the
supersymmetric standard model the gauge coupling constants of the
strong, weak and electromagnetic interactions, with SUSY broken at the
TeV-scale, meet at a single point $M_{\rm GUT} \simeq (2-3) \times
10^{16}$ GeV.  These are called Supersymmetric Grand Unified Theories
(SUSY GUTs).  An acceptable SUSY GUT model should be consistent with
the standard model as well as with cosmology.  SUSY GUTs can provide
the scalar field needed for inflation, they can explain the
matter-antimatter asymmetry of the Universe, and they can provide a
candidate for cold dark matter, known as the lightest superparticle.

Usually models of SUSY GUTs suffer from the appearance of undesirable
stable topological defects, which are mainly monopoles, but also
domain walls.  Topological defects appear via the Kibble
mechanism~\cite{kibble}. A common mechanism to get rid of the unwanted
topological defects, is to introduce one or more inflationary stages.
Inflation essentially consists of a phase of accelerated expansion
which took place at a very high energy scale.  In addition, inflation
provides a natural explanation for the origin of the large scale
structure and the associated temperature anisotropies in the Cosmic
Microwave Background (CMB) radiation. On the other hand, one has to
address the question of whether it is needed fine-tuning of the
parameters of the inflationary model. This is indeed the case in some
nonsupersymmetric versions of inflation and it leads to the
naturalness issue. Even though SUSY models of hybrid inflation were for 
long believed to circumvent these fine-tuning issues, our study
shows that this may not be the case, unless we invoke the curvaton mechanism 
for the origin of the initial density fluctuations.

The study we present here is the continuation of Ref.~\cite{rjs}, where they
were examined all possible Spontaneously Symmetry Breaking (SSB)
schemes from a large gauge group down to the SM gauge group, in the
context of SUSY GUTs. Assuming standard supersymmetric hybrid
inflation, there were found all models which are consistent with high energy
physics and cosmology. Namely, there were selected all models which can solve
the GUT monopole problem, lead to baryogenesis after inflation and are
consistent with proton lifetime measurements. That study led  to the
conclusion that in all acceptable SSB patterns, the formation of 
cosmic strings is unavoidable, and some times it is accompanied by the
formation of embedded strings. The strings which form at the end of
inflation have a mass which is proportional to the inflationary scale.
Here, we find the inflationary scale, which coincides with the string mass
scale. Since our analysis is within global supersymmetry, we examine 
whether the value of the inflaton is at least a few orders of magnitude 
smaller than the Planck scale. As we show, global supersymmetry is
sufficient in the case of F-term inflation, while D-term inflation
necessitates the supergravity framework.

We organise the rest of the paper as follows: In Section II, we
discuss the theoretical framework of our study. We briefly review the
choice of the gauge groups we consider and we state the results about
SSB schemes allowed from particle physics and cosmology as well as the
topological defects left after the last inflationary era. We briefly
review the theory of CMB temperature fluctuations. We then discuss
inflation within N=1 SUSY GUTs, first in the context of F-term and
then in the context of D-term inflation. In Section III, we describe
our analysis as well as our findings for the mass scale and the
strings contribution to the CMB. We first discuss F-term and then that
of D-term inflation. We show that F-term inflation can be addressed in
the context of global supersymmetry. We then show that D-term
inflation has to be studied within local supersymmetry and we give the
scalar potential for D-term inflation, taking into account radiative
and supergravity (SUGRA) corrections.  We round up with our
conclusions in Section IV. In Appendix A, we list the allowed
SSB schemes of large gauge groups down to the standard model, found
in Ref.~\cite{rjs}, which
are allowed by particle physics and cosmology.

\section{Theoretical Framework}
\subsection{Hybrid inflation within supersymmetric grand unified theories}

Grand Unified Theories imply a sequence of phase transitions
associated with the SSB of the GUT gauge group ${\rm G}_{\rm GUT}$
down to the standard model gauge group ${\rm G}_{\rm SM} = {\rm
SU}(3)_{\rm C} \times {\rm SU}(2)_{\rm L} \times {\rm U}(1)_{\rm Y}$. The 
energy scale at which this sequence of SSBs starts is $M_{\rm GUT} 
\sim 3 \times 10^{16}$ GeV. As our Universe has
undergone this series of phase transitions, various kinds of
topological defects may have been left behind as the consequence of
SSB schemes, via the Kibble mechanism~\cite{kibble}.  Among the
various kinds of stable topological defects, monopoles and domain
walls are undesirable, since they lead to catastrophic cosmological
implications, while textures do not have important cosmological 
consequences. To get rid of the
unwanted topological defects one may employ the mechanism of cosmological
inflation.

Considering supersymmetric grand unified theories is motivated 
by several constraints, coming from both particle physics and cosmology.
It is indeed the only viable framework to solve the hierarchy problem.
In addition, it allows for a unification of strong, weak and electromagnetic 
interactions at a sufficiently high scale to be 
compatible with proton lifetime measurements. From the point of view of 
cosmology, SUSY GUTs can provide a good candidate for dark matter, namely the
lightest supersymmetric particle. Moreover, it offers various 
candidates for playing the r\^ole of the inflaton field and it can give naturally 
a flat direction for slow-roll inflation.

Supersymmetry can be formulated either as a global or as a local
symmetry, in which case gravity is included and the theory is called
supergravity. Global supersymmetry can be seen as a limit of
supergravity and it is a good approximation provided the Vacuum Expectation 
Values (VEVs) of all relevant fields are much smaller than the Planck mass.

In a supersymmetric theory, the tree-level potential for a scalar field is the
sum of an F-term and a D-term. These two terms have rather different
properties and in all proposed hybrid inflationary models only one of the two
terms dominates.  In supersymmetric hybrid inflation, the superpotential
couples an inflaton field to a pair of Higgs fields that are responsible for
one symmetry breaking in the SSB scheme.  This class of inflationary models is
considered natural~\cite{RJ1} within SUSY GUTs in the sense that the only
extra field which is added, except the fields needed to build the GUT itself,
is a singlet scalar field. This extra field is however likely to be anyway
needed to build the GUT model, so that it constraints the Higgs fields to
acquire a VEV. Moreover, it is not spoiled by radiative corrections and
supergravity corrections can be kept small for F-term inflation. We would 
like to mention that generically, F-term inflation suffers from the 
so-called $\eta$-problem, because usually, the supergravity corrections induce
contribution of order unity to the slow roll parameter $\eta\equiv M_{\rm Pl}^2 
V''/V$. This effective mass term for the inflaton field would spoil the slow roll 
condition~\cite{sugra}. 
However, in the case of a minimal Kahler potential, which is what is considered 
hereafter, this problem is lifted by a cancelation of the problematic mass terms. 
This can however be seen as a fine tunning since a minimal kahler potential is 
not well motivated for example from the point of view of the string theory~\cite{sugra}. 
Finally it
has been stated that such inflationary models are successful in the sense that
they do not require any fine tuning, but we show that this last point has to be 
revisited. 

F-term inflation can occur naturally within the GUTs framework when for
example, a GUT gauge group $G_{\rm GUT}$ is broken down to the SM 
at an energy $M_{\rm GUT}$ according to
\begin{equation}
G_{\rm GUT} \stackrel{M_{\rm GUT}}{\hbox to 0.8cm {\rightarrowfill}} H_1
\xrightarrow{9}{M_{\rm infl}}{1}{\Phi_+\Phi_-} H_2 {\longrightarrow} 
G_{\rm SM}~,
\end{equation}
where $\Phi_+, \Phi_-$ is a pair of GUT Higgs superfields in
nontrivial complex conjugate representations, which lower the rank of
the group by one unit when acquiring nonzero VEV. The inflationary phase takes
place at the beginning of the symmetry breaking 
$H_1\stackrel{M_{\rm infl}}{\longrightarrow} H_2$.

F-term inflation is based on the globally supersymmetric renormalisable 
superpotential
\begin{equation}\label{superpot}
W_{\rm infl}^{\rm F}=\kappa S(\Phi_+\Phi_- - M^2)~,
\end{equation}
where $S$ is a GUT gauge singlet left handed superfield and $\Phi_+,
\Phi_-$ are as defined above, with $\kappa$ and
$M$ two constants ($M$ has dimensions of mass) which can both be taken
positive with field redefinition.  The chiral superfields $S, \Phi_+,
\Phi_-$ are taken to have canonical kinetic terms.  The above superpotential
 is the most general superpotential
consistent with an R-symmetry under which $W \rightarrow e^{i \beta}
W~, \Phi_- \rightarrow e^{-i \beta} \Phi_-~, \Phi_+ \rightarrow e^{i
\beta} \Phi_+$, and $S \rightarrow e^{i \beta} S$. We note that an
R-symmetry can ensure that the rest of the renormalisable terms are
either absent or irrelevant.

D-term inflation is derived from the superpotential
\begin{equation}
\label{superpotD}
W^{\rm D}_{\rm infl}=\lambda S \Phi_+\Phi_-~,
\end{equation}
where $S, \Phi_-, \Phi_+$ are three chiral superfields and $\lambda$ is the
superpotential coupling.  D-term inflation requires the existence of a nonzero
Fayet-Iliopoulos term $\xi$, which can be added to the lagrangian only in the
presence of a U(1) gauge symmetry, under which, the three chiral superfields
have charges $Q_S=0$, $Q_{\Phi_+}=+1$ and $Q_{\Phi_-}=-1$, respectively.
Thus, D-term inflation requires a scheme, like
\begin{equation}
G_{\rm GUT}\times U(1) \stackrel{M_{\rm GUT}}{\hbox to 0.8cm{\rightarrowfill}} H \times U(1)
\xrightarrow{9}{M_{\rm infl}}{1}{\Phi_+\Phi_-}
H \rightarrow G_{\rm SM}~.
\end{equation}
This extra U(1) gauge symmetry symmetry can be of a different
origin~\cite{RJ1}. In what follows, we consider a nonanomalous U(1) gauge
symmetry. We note however that one could instead consider a situation realised
in heterotic string theories, where there is an anomalous D-term arising from
an anomalous U(1)$_{\rm A}$, which can contribute to the vacuum
energy~\cite{anomalousU1}. Clearly, the symmetry breaking at the end of the
inflationnary phase implies that cosmic strings are always formed at the end
of D-term hybrid inflation.

In the SSB schemes studied in Ref.~\cite{rjs}, one can naturally
incorporate an era of F-/D-term inflation.  All SSB schemes from grand unified
gauge groups ${\rm G_{GUT}}$ of rank at the most equal to 8 down to the
standard model gauge group ${\rm G_{SM}}\times Z_2$ have been considered.
Initially, the group ${\rm G_{GUT}}$ was chosen to be one of the following
ones: SU(5), SO(10), SU(6), E$_6$, SU(7), SO(14), SU(8), and SU(9). In
addition, the choice of ${\rm G_{GUT}}$ was limited to simple gauge groups
which contain ${\rm G_{SM}}$, have a complex representation, are anomaly free
and take into account some of the major observationnal constraints of particle
physics and cosmology. The above $Z_2$ symmetry is a sub-group of the ${\rm U}
(1)_{\rm B-L}$ gauge symmetry which is contained in various gauge groups and it 
plays the r\^ole of R-parity. R-parity can only appear in grand unified gauge 
groups which contain ${\rm U}(1)_{\rm B-L}$.  There were considered as many 
possible embeddings of ${\rm G_{SM}}$ in ${\rm G_{GUT}}$ as one can find in the
literature and it was examined whether defects are formed during the SSB
patterns of ${\rm G_{GUT}}$ down to ${\rm G_{SM}}$, and of which kind they
are. Assuming standard hybrid F-term inflation there were disregarded all SSB
patterns where monopoles or domain walls are formed after the end of the last
possible inflationary era. In addition, there were disregarded SSB schemes
with broken R-parity since the proton would decay too rapidly as compared to
Super-Kamiokande measurements. It was also required that the gauged U(1)$_{\rm
B-L}$ symmetry, is broken at the end of inflation so that a non thermal
leptogenesis can explain the baryon/antibaryon asymmetry in the
Universe. SU(5), SU(6), SU(7) are thus not acceptable groups for particle
physics, since SU(5) leads to the formation of stable monopoles, while minimal
SU(6) and minimal SU(7) do not contain U(1)$_{\rm B-L}$.

The SSB schemes compatible with high energy physics and cosmology, as
given in detail in Ref.~\cite{rjs}, are listed in Appendix A.  

It was concluded that within the framework of the analysis of Ref.~\cite{rjs},
there are not any acceptable SSB schemes without cosmic strings after the last
inflationary era. The analysis we present below can therefore be applied to
all these models.  We would like to note that even if we relax the requirement
that the gauged ${\rm B-L}$ symmetry is broken at the end of inflation,
the results of Ref.~\cite{rjs} remain unchanged.

Even if one also allows for patterns with broken R-parity at low energy,
cosmic strings formation is still very generic. To consider these cases, one
should however find a mechanism to protect proton lifetime from very dangerous
dimension 4 operators.  Strings formed during the SSB phase transitions
leading from a large gauge group ${\rm G_{GUT}}$ down to the G$_{\rm SM}\times
Z_2$ are of two types: topological strings, called cosmic strings, and
embedded strings which are not topologically stable and in general they are
not dynamically stable either.

Clearly, as we discussed earlier, in the case of D-term inflation, cosmic
strings are always present at the end of the inflationary era.

\subsection{Cosmic microwave background temperature anisotropies}

The CMB temperature anisotropies provide a powerful test for
theoretical models aiming at describing the early Universe.  The
characteristics of the CMB  multipole moments, and more
precisely the position and amplitude of the acoustic peaks, as well as
the statistical properties of the CMB temperature anisotropies, can be used to
discriminate among theoretical models, as well as to constrain the
parameters space.

The spherical harmonic expansion of the cosmic microwave background
temperature anisotropies, as a function of angular position, is given by
\begin{equation}
\label{dTT}
\frac{\delta T}{T}({\bf n})=\sum _{\ell m}a_{\ell m} {\cal W}_\ell
Y_{\ell m}({\bf n})~\,
\end{equation}
 with
\begin{equation}
a_{\ell m}=\int {\rm
d}\Omega _{{\bf n}}\frac{\delta T}{T}({\bf n})Y_{\ell m}^*({\bf n})~,
\end{equation}
where ${\cal W}_\ell $ stands for the $\ell$-dependent window
function of the particular experiment.

The angular power spectrum of CMB anisotropies is expressed in terms
of the dimensionless coefficients $C_\ell$, which appear in the
expansion of the angular  correlation function in terms of the
Legendre polynomials $P_\ell$:
\begin{equation}
\biggl \langle 0\biggl |{\delta T\over T}({\bf n}){\delta T\over
T}({\bf n}') \biggr |0\biggr\rangle \left|_{{~}_{\!\!({\bf n\cdot
n}'=\cos\vartheta)}}\right. = {1\over 4\pi}\sum_\ell(2\ell+1)C_\ell
P_\ell(\cos\vartheta) {\cal W}_\ell^2 ~.
\label{dtovertvs}
\end{equation}
It compares points in the sky separated by an angle $\vartheta$.
Here, the brackets denote spatial average, or expectation values if
perturbations are quantised~\footnote{We emphasise that Eq.~(\ref{dtovertvs})
holds only if the initial state for cosmological perturbations of 
quantum-mechanical origin is the vacuum~\cite{jm1,jm2}.}.  The value of 
$C_\ell$ is determined by
fluctuations on angular scales of order $\pi/\ell$. The angular power
spectrum of anisotropies observed today is usually given by the power
per logarithmic interval in $\ell$, plotting $\ell(\ell+1)C_\ell$
versus $\ell$.  The coefficients $C_\ell$ are related to $a_{\ell m}$
by
\begin{equation}
C_\ell = {\langle\sum_m | a_{\ell m} |^2 \rangle \over 2\ell+1}~.
\end{equation}

On large angular scales, the main contribution to the temperature
anisotropies is given by the Sachs-Wolfe effect, implying
\begin{equation}
\label{sw}
\frac{\delta T}{T}({\bf n})\simeq
\frac{1}{3}\Phi [\eta _{\rm lss},{\bf n}(\eta _0-\eta _{\rm lss})],
\end{equation}
where $\Phi (\eta ,{\bf x})$ is the Bardeen potential, while $\eta _0$
and $\eta _{\rm lss}$ denote respectively the conformal times now and
at the last scattering surface. Note that the previous expression is
only valid for the standard cold dark matter model.

If we assume that the initial density perturbations are due to
``freezing in'' of quantum fluctuations of a scalar field during an
inflationary period, then the quadrupole anisotropy reads
\begin{equation}\label{contribInfl}
\left(\frac{\delta T}{T}\right)_{\rm Q-infl} =
\left[\left(\frac{\delta T}{T}\right)_{\rm Q- scal}^2 +
\left(\frac{\delta T}{T}\right)_{\rm Q- tens}^2\right]^{1/2}~,
\end{equation}
where the quadrupole anisotropy due to the scalar Sachs-Wolfe effect
is 
\begin{equation}\label{contribInflScal}
\left(\frac{\delta T}{T}\right)_{\rm Q- scal} =
\frac{1}{4\sqrt{45}\pi}\frac{V^{3/2}(\varphi_Q)}{M_{\rm
Pl}^3\,V'(\varphi_Q)}~,
\end{equation}
and the tensor quadrupole anisotropy is
\begin{equation}
\label{contribInflTens}
\left(\frac{\delta T}{T}\right)_{\rm Q-tens}\sim {0.77\over 8\pi}
\,\frac{V^{1/2}(\varphi_Q)}{M_{\rm Pl}^2}.
\end{equation}
We note that $V$ is the potential of the inflaton field $\varphi$,
 with $V'\equiv {\rm d}V(\varphi)/{\rm d}\varphi$, while $M_{\rm Pl}$
 denotes the reduced Planck mass, $M_{\rm Pl}= {(8\pi
 G)^{-1/2}}\simeq 2.43\times 10^{18}$ GeV, and $\varphi_{\rm
 Q}$ is the value of the inflaton field when the comoving
 scale corresponding to the quadrupole anisotropy became 
bigger than the Hubble radius.

The number of e-foldings of inflation between the initial value of 
inflaton $\varphi_i$ and the final value $\varphi_f$ is given by
\begin{equation}
\label{defnefold}
N(\varphi_i \rightarrow \varphi_f)=-8\pi G
\int_{\varphi_i}^{\varphi_f}{V(\varphi) \over V'(\varphi)}{\rm
d}\varphi~.
\end{equation}

The primordial fluctuations could also be generated from the quantum
fluctuations of a late-decaying scalar field other than the inflaton,
known as the curvaton field ${\cal \psi}$~\cite{lw2002,mt2001,ekm2003,dl2004},
whose
nonvanishing initial amplitude is denoted by ${\cal \psi}_{\rm
init}$.  During inflation the curvaton potential is very flat and
${\cal \psi}$ acquires quantum fluctuations~\cite{mt2001}
\begin{equation}
\delta{\cal \psi}_{\rm init}={H_{\rm inf}\over 2\pi}~,
\label{qfc}
\end{equation}
where $H_{\rm infl}$ denotes the expansion rate during inflation,
which is a function of the inflaton field, and it is given by the Friedmann 
equation
\begin{equation}
H_{\rm infl}^2(\varphi)={8\pi G\over 3} V(\varphi)~.
\label{exprate}
\end{equation}
We  assumed that the effective curvaton mass is much
smaller than $H_{\rm infl}$, since otherwise the quantum fluctuations
of the curvaton field during inflation would be very small and the CMB
power spectrum would remain the same as in the standard adiabatic case.

At the end of the inflationary era, $\delta{\cal \psi}_{\rm init}$
generates an entropy fluctuation. At later times, the curvaton field
first dominates the energy density of the Universe, and then (during
the RDE) it decays and reheats the Universe. Since the primordial
fluctuations of the curvaton field are converted to purely adiabatic
density fluctuations, the effect of $\delta{\cal \psi}_{\rm init}$ in
the CMB angular power spectrum, which can be parametrised by the
metric perturbation induced by the curvaton fluctuation, is given
by~\cite{mt2002}
\begin{equation}
\Psi_{\rm curv}=-{4\over 9}{\delta{\cal \psi_{\rm init}}\over
\psi_{\rm init}}~.
\label{Psi_curv}
\end{equation}
There is no correlation between the primordial fluctuations of the
inflaton and curvaton fields.

As we have explicitely shown in Ref.~\cite{rjs}, the end of the
inflationary era is accompanied by strings formation~\footnote{Some
authors have proposed mechanisms which could avoid cosmic strings
production at the end of  hybrid inflation. This is for example realised by
adding a nonrenormalisable term in the superpotential~\cite{shifted},
or by adding an additional discrete symmetry~\cite{smooth}, or by
considering GUTs models based on nonsimple and more complicated
groups~\cite{watari}. Moreover, by introducing a new pair of charged 
superfields in the framework of an N=2 supersting version of D-term 
inflation the strings which form are nontopological~\cite{davis}. 
This last model was shown~\cite{davis} to satisfy the CMB constraints.}, 
which are cosmic strings (topological defects), sometimes accompagnied 
by embedded strings (not topologically 
stable and in general not dynamically stable either).  Let us first
calculate the contribution to the quadrupole temperature anisotropy
coming from the cosmic strings network.

At this point, we would like to bring to the attention of the reader
that since the calculation of the angular power spectrum induced by a
cosmic strings network relies on heavy numerical simulations, this
issue remains still open.  More precisely, to obtain the power
spectrum from numerical simulations with cosmic strings, one must take
into account the ``three-scale model''~\cite{3scale} of cosmic strings
networks, the small-scale structure (wiggliness) of the strings, the
microphysics of the strings network, as well as the expansion of the
Universe.  This is indeed a rather difficult task, and to our
knowledge no currently available simulation leading to the
$C_\ell^{\rm strings}$ includes all of them.  Moreover, all
simulations of cosmic strings are referred to Nambu-Goto strings, and
this is probably the most unrealistic case~\footnote{The nature of
cosmic strings formed in the context of our models within SUSY GUTs and their
cosmological r\^ole is studied in Ref.~\cite{jm4}. One has to examine whether 
the fermion zero modes which are intrinsic in supersymmetric models of cosmic 
strings lead to the production of vortons, which may result to a cosmological
problem~\cite{davis2}. The microphysics of cosmic string solutions to N=1 
supersymmetric abelian Higgs models has been studied in Ref.~\cite{davis3}.}
At least, but not last, there is the issue of the appropriate initial 
conditions for the time evolution of the cosmic strings network. More 
precisely,
we do not know of any numerical simulation leading to the $C_\ell^{\rm
strings}$, where the initial configuration of the strings network was
other than the one obtained by assigning at random values to a phase
variable on a cubic lattice. One should probably study whether the
strings network is in the low or high density regime~\footnote{When
the energy density of the strings network is low, the dominant part of
the strings is in the form of closed loops of the smallest allowed
size. At a certain critical density the strings network undergoes a
phase transition characterised by the formation of {\sl
infinite} strings~\cite{sv,s}.}.
 
Nevertheless, in what follows we use recent results~\cite{ls2003}
based on Nambu-Goto local strings simulations in a
Friedmann--Lema\^{\i}tre--Roberston--Walker spacetime and we assume
\begin{equation}\label{contribCS}
\left(\frac{\delta T}{T}\right)_{\rm cs}\sim (9-10) G\mu \quad
\mathrm{with}\quad \mu=2\pi\langle\chi \rangle^2~,
\end{equation}
where $\langle\chi \rangle$ is the Vacuum Expectation Value (VEV) of
the Higgs field responsible for the formation of cosmic strings.

\subsection{Inflation in supersymmetric grand unified theories}

 In what follows, we first discuss inflation where
the F-term dominates (F-term inflation) and then we address inflation
where the D-term dominates (D-term inflation).

\subsubsection{F-term inflation}

F-term inflation is based on the globally supersymmetric
renormalisable superpotential  Eq.~(\ref{superpot}).
The scalar potential $V$ can be obtained from Eq.~(\ref{superpot}) and
it reads
\begin{equation}
\label{scalpot1}
V(\phi_+,\phi_-,S)=
|F_{\Phi_+}|^2+|F_{\Phi_-}|^2+|F_S|^2 +{1\over 2}\sum_a g_a^2 D_a^2~,
\end{equation}
where the F-term is such that\footnote{The notation $|\partial
W/\partial \Phi_i|_{\theta=0}$ means that one has to take the scalar
component (with $\theta =\bar{\theta}=0$ in the superspace) of the
superfields once one differentiates with respect to the superfields
$\Phi_i$.  This is the reason for which the potential $V$ is a function of the scalar
fields $\phi_i$.} $F_{\Phi_i} \equiv |\partial W/\partial
\Phi_i|_{\theta=0}$, with $\Phi_i=\Phi_+, \Phi_-, S$, and
\begin{equation}
D_a=\bar{\phi}_i\,{(T_a)^i}_j\,\phi^j +\xi_a~,
\end{equation}
with $a$ the label of the gauge group generators $T_a$, $g_a$ the
gauge coupling, and $\xi_a$ the Fayet-Iliopoulos term. By
definition, in the F-term inflation the real constant $\xi_a$ is zero;
it can only be nonzero if $T_a$ generates a U(1) group.

In the case of F-term inflation, the potential $V$ as a function
of the complex scalar component of the respective chiral superfields
$\Phi_+,\Phi_-, S$ reads
\begin{equation}
\kern -2cm V^{\rm F}(\phi_+,\phi_-,S)=\kappa^2|M^2-\phi_-\phi_+|^2
+\kappa^2|S|^2(|\phi_-|^2+|\phi_+|^2)
+\mathrm{D-terms}.
\end{equation}
Assuming that the F-terms give rise to the inflationary potential
energy density, while the D-terms are flat along the inflationary
trajectory,  one may neglect them during inflation. 
The D-terms may play an important r\^ole in determining the
trajectory and in stabilising the noninflaton fields.

The potential has two minima: one valley of local minima ($V=\kappa^2 M^4
\equiv V_0$), for $S$ greater than its critical value $S_{\rm c} = M $ with
$\phi_+ = \phi_-=0$, and one global supersymmetric minimum ($V=0$) at $S=0$
and $\phi_+ = \phi_- = M$. Imposing chaotic initial conditions, i.e. $ S \gg
S_{\rm c}$, the fields quickly settle down the valley of local minima.  In the
slow roll inflationary valley ($\phi_+=\phi_-=0$, $|S|\gg M$), the ground
state of the scalar potential is different from zero, meaning that SUSY is
broken.  In the tree level, along the inflationary valley the potential
$V=V_0$ is constant and thus it is perfectly flat. A slope along the potential
can be generated by including the one-loop radiative corrections which are
small as compared to $V_0$. Thus, the scalar potential acquires a little tilt
which helps the scalar field $S$ to slowly roll down the valley of minima. The
SSB of SUSY along the inflationary valley by the vacuum energy density
$\kappa^2M^4$ leads to a mass splitting in the superfields $\Phi_+,
\Phi_-$. One gets \cite{DvaShaScha} a Dirac fermion with a mass squared term
$\kappa^2|S|^2$ and two complex scalars with mass squared terms
$\kappa^2|S|^2\pm \kappa^2M^2$. This implies that there are one-loop radiative
corrections to $V$ along the inflationary valley, which can be calculated
using\footnote{This expression has been derived in the case of a Minkowski
background. However, during inflation the background geometry is given by the
De Sitter metric, and therefore, strictly speaking, one should not use the
standard Coleman-Weinberg expression, but one should instead find the
corresponding expression in a De Sitter background.} the Coleman-Weinberg
expression~\cite{cw}
\begin{equation}\label{cw}
\Delta V_{1-{\rm loop}}=\frac{1}{64\pi^2}\sum_i (-1)^{F_i}
m_i^4\ln\frac{m_i^2}{\Lambda^2}~,
\end{equation}
where the sum extends over all helicity states i, with fermion number
$F_i$ and mass squared $m_i^2$; $\Lambda$ stands for a
renormalisation scale. Thus, the effective potential
reads~\cite{DvaShaScha,lr,dm,Lazarides,SenoSha}
$$\kern -8cm V_{\rm eff}^{\rm F}(|S|)=V_0+[\Delta
V(|S|)]_{1-{\rm loop}}$$
\begin{equation}
\kern -1.2cm =\kappa^2M^4\biggl\{1+\frac{\kappa^2
\cal{N}}{32\pi^2}\biggl[2\ln\frac{|S|^2\kappa^2}{\Lambda^2}+
(z+1)^2\ln(1+z^{-1})
+(z-1)^2\ln(1-z^{-1})\biggl]\biggl\}
~,
\label{VexactF}
\end{equation}
where
\begin{equation}
z=\frac{|S|^2}{M^2}\equiv x^2~,
\end{equation}
and $\cal{N}$ stands for the dimensionality of the representations to
which the fields $\phi_+, \phi_-$ belong.

Employing the above found expression, Eq.~(\ref{VexactF}), for the
effective potential we calculate in the next section the inflaton
and cosmic strings contributions to the CMB temperature anisotropies.

\subsubsection{D-term inflation}

In the context of global supersymmetry, D-term inflation is derived
from the superpotential  Eq.~(\ref{superpotD}).
In the global supersymmetric limit, Eqs.(\ref{scalpot1}),~(\ref{superpotD})
 lead to the following expression for the scalar
potential
\begin{equation}
\label{VtotD}
\kern -2cm V^{\rm D}(\phi_+,\phi_-,S) = \lambda^2
\left[\,|S|^2(|\phi_+|^2+|\phi_-|^2) + |\phi_+\phi_-|^2 \right]
+\frac{g^2}{2}(|\phi_+|^2-|\phi_-|^2+\xi)^2~,
\end{equation}
where $g$ is the gauge coupling of the U(1) symmetry and $\xi$ is a
Fayet-Iliopoulos term, chosen to be positive.

The potential has two minima. There is one global minimum at zero,
reached for $|S|=|\phi_+|=0$ and $|\phi_-|=\sqrt{\xi}$. There is also
one local minimum, found by minimising the potential for fixed values
of $S$ with respect to the other fields. This local minimum is equal
to $V_0=g^2\xi^2/2$, reached for $|\phi_+|=|\phi_-|=0$, with
$|S|>S_{\rm c}\equiv g\sqrt{\xi}/\lambda$. As in the previous
discussed case (F-term inflation), also here the SSB of supersymmetry
in the inflationary valley introduces a splitting in the masses of the
components of the chiral superfields $\Phi_+$ and $\Phi_-$. As a
result, we obtain two scalars with squared masses $m^2_{\pm}=\lambda^2
|S|^2 \pm g^2\xi$ and a Dirac fermion with squared mass $\lambda^2 |S|^2$.

For arbitrary large $S$ the tree level value of the potential remains
constant and equal to $V_0=(g^2/2)\xi^2$, thus $S$ plays naturally the
r\^ole of an inflaton field. Assuming chaotic initial conditions
$|S|\gg S_{\rm c}$ one can see the onset of inflation. Along the
inflationary trajectory the F-term vanishes and the Universe is
dominated by the D-term, which splits the masses in the $\Phi_+$ and
$\Phi_-$ superfields, resulting to the one-loop effective potential
for the inflaton field.

The radiative corrections given by the Coleman-Weinberg  expression
Eq.~(\ref{cw}) lead to the following effective potential for D-term
inflation
$$\kern -7.8cm V^{\rm D}_{\rm eff}(|S|) = V_0 + [\Delta V(|S|)]_{1-{\rm
loop}}$$
\begin{equation}
\kern -1cm = \frac{g^2\xi^2}{2}\biggl\{1+\frac{g^2}{16\pi^2}
\biggl[2\ln\frac{|S|^2\lambda^2}{\Lambda^2}+
(z+1)^2\ln(1+z^{-1})
+(z-1)^2\ln(1-z^{-1})\biggl]
\biggl\}~,
\label{VexactD}
\end{equation}
with
\begin{equation}\label{xdeS}
z=\frac{\lambda^2 |S|^2}{g^2\xi}\equiv x^2~.
\end{equation}

Employing the above found expression for the effective potential,
Eq.~(\ref{VexactD}), we calculate in the next section the inflaton and
cosmic strings contributions to the CMB temperature anisotropies.

\section{Energy scale of inflation and inflaton/cosmic strings 
contributions to the CMB}

\subsection{F-term inflation in global supersymmetry}

Assuming $V\simeq \kappa^2M^4$, while using the exact expression for
the potential as given in Eq.~(\ref{VexactF}) for
calculating $V'\equiv {\rm d}V/{\rm d}S$, we obtain
\begin{equation}
V'(|S|)=\frac{2a}{|S|}z\,f(z)~,
\end{equation}
with
\begin{equation}
a=\frac{\kappa^4 M^4 \mathcal{N}}{16 \pi^2}~,
\end{equation}
and
\begin{equation}\label{fdez}
f(z)=(z+1)\ln(1+z^{-1})+(z-1)\ln(1-z^{-1})~.
\end{equation}
Equation (\ref{contribInflScal}) implies~\cite{Lazarides,SenoSha} 
\begin{equation}\label{contribInflScal2}
\left(\frac{\delta T}{T}\right)_{\rm Q-scal} \sim
\frac{1}{\sqrt{45}}\,\sqrt{\frac{N_{\rm Q}}{\cal N}}\,
\frac{M^2}{M_{\rm Pl}^2}\,
x_{\rm Q}^{-1}y_{\rm Q}^{-1}f^{-1}(x_{\rm Q}^2)~,
\end{equation}
and Eq.~(\ref{defnefold}) leads to 
\begin{equation}
N_Q=\frac{4\pi^2}{\kappa^2\cal N}\frac{M^2}{M_{\rm Pl}^2}\,y_{\rm Q}^2~,
\end{equation}
with
\begin{equation}
\label{yQ}
y_{\rm Q}^2=\int_1^{x_{\rm Q}^2}\frac{{\rm d}z}{zf(z)}~,
\end{equation}
and 
\begin{equation}
x_{\rm Q}={|S_{\rm Q}|\over M}~.
\end{equation}
We remind to the reader that the index $_{\rm Q}$ denotes the scale
responsible for the quadrupole anisotropy in the CMB.

The coupling $\kappa$ is related to the mass scale $M$, through the
relation 
\begin{equation}\label{Mdekappa}
\frac{M}{M_{\rm Pl}}=\frac{\sqrt{N_{\rm Q} \cal N}\,\kappa}{2\,
\pi\,y_{\rm Q}}~.
\end{equation}
Equation (\ref{contribInflTens}) implies~\cite{Lazarides,SenoSha} 
\begin{equation}\label{contribInflTens2}
\left(\frac{\delta T}{T}\right)_{\rm Q-tens} \sim
{0.77\over 8\pi} \kappa {M^2\over M_{\rm Pl}^2}~.
\end{equation}
Employing Eq.~(\ref{contribCS}) in the case of F-term inflation, we
express the cosmic strings contribution as
\begin{equation}
\left(\frac{\delta T}{T}\right)_{\rm cs}\sim
{9\over 4}\left(\frac{M}{M_{\rm Pl}} \right)^2~.
\end{equation}
Therefore, the total quadrupole anisotropy
\begin{equation}\label{dttot}
\left[\left(\frac{\delta T}{T}\right)_{\rm Q-tot}\right]^2=
\left[\left(\frac{\delta T}{T}\right)_{\rm
scal}\right]^2+\left[\left(\frac{\delta T}{T}\right)_{\rm
tens}\right]^2+\left[\left(\frac{\delta T}{T}\right)_{\rm cs}\right]^2
\end{equation}
is explicitely given by
\begin{equation}
\label{eqnumF}
\kern -2.5cm\left({\delta T\over T}\right)_{\rm Q-tot} \sim\Big\{y_{\rm
Q}^{-4}\left({\kappa^2 \mathcal{N}\, N_{\rm Q}\over 32\pi^2}\right)^2
\Big[\frac{64N_{\rm Q}}{45\cal N} x_{\rm Q}^{-2}y_{\rm
Q}^{-2}f^{-2}(x_Q^2)
+\left(\frac{0.77 \kappa}{\pi}\right)^2 +
324\Big]\Big\}^{1/2}~,
\end{equation}
where the $\left(\delta T/ T\right)_{\rm Q}^{\rm tot}$ is normalised
to the Cosmic Background Explore (COBE) data~\cite{cobe}, namely
$\left(\delta T/ T\right)_{\rm Q}^{\rm COBE} \sim 6.3\times 10^{-6}$.
For given values of $\kappa, N_{\rm Q}, \mathcal{N}$, the above
equation can be solved numerically for $x_{\rm Q}$, and then employing
Eqs.~({\ref{yQ}) and (\ref{Mdekappa}), one obtains $y_{\rm Q}$ and
$M$.  We assume $N_{\rm Q}=60$ and we find the inflationary scale $M$
which is proportional to the string mass scale, as a function of the
superpotential coupling $\kappa$ for three values of $\mathcal{N}$.
We choose $\mathcal{N}=\mathbf{27}, \mathbf{126}, \mathbf{351}$, which 
correspond to realistic SSB schemes in SO(10) or E$_6$ models. The
results are shown in Fig.~\ref{inflscaleF} below.

\begin{figure}[hhh]
\begin{center}
\includegraphics[scale=.6]{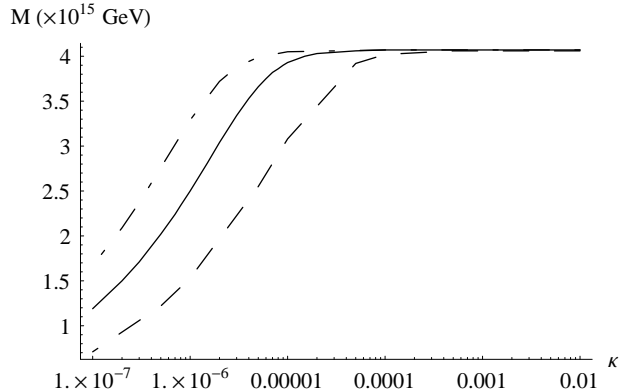}
\caption{Evolution of the inflationary scale $M$ in units of $10^{15}$ GeV as
a function of the dimensionless coupling $\kappa$.  The three curves
correspond to $\mathcal{N}=\mathbf{27}$ (curve with broken line),
$\mathcal{N}=\mathbf{126}$ (full line) and $\mathcal{N}=\mathbf{351}$ (curve
with lines and dots).  }
\label{inflscaleF}
\end{center}
\end{figure}

Clearly, the mass scale is of the order of $10^{15}$
GeV, and it grows very slowly with $\mathcal{N}$. Since it will be
useful later to know approximately the evolution of the mass parameter
$M$ with respect to the coupling  $\kappa$, we fit the curve
for $\mathcal{N}=\mathbf{126}$, which is shown in Fig.~\ref{inflscaleF}, by
\begin{equation}\label{fitM}
\kern -2cm M(\kappa)\sim \left\{
  \begin{array}{ll}
    4.1\times 10^{15} + 5.3\times 10^{12}\,\ln(\kappa)~{\rm GeV}~ &
    ~\mathrm{if} \quad\kappa \in [2\times 10^{-5},10^{-2}]~, \\ 1.1\times
    10^{16} + 6.3\times 10^{14}\,\ln(\kappa)~{\rm GeV}~ & ~\mathrm{if}
    \quad\kappa \in [10^{-7},2\times 10^{-5}]~. \\
  \end{array}
  \right.
\end{equation}

The dimensionless coupling  $\kappa$ is subject to the
gravitino constraint, as well as to the constraint imposed by the
CMB temperature anisotropies. As we show below, the strongest
constraint on $\kappa$ is imposed by the measured CMB temperature
anisotropies; the cosmic strings contribution to the power spectrum
should be strongly suppressed in order not to contradict the data.

Firstly, we calculate the upper bound on the dimensionless superpotential
coupling  $\kappa$, as imposed by the gravitino constraint.
After the inflationary era, our Universe enters the high entropy
radiation dominated phase via the reheating process, during which the
inflaton energy density decays perturbatively into normal
particles. This process is characterised by the reheating temperature
$T_{\rm RH}$.  In order to have a successful reheating it is important
not to create too many gravitinos.

Within the Minimal Supersymmetric Standard Model (MSSM), and assuming
a see-saw mechanism to give rise to massive neutrinos, the reheating
temperature is~\cite{laz} 
\begin{equation}
\label{rhn}
T_{\rm RH}\approx {(8\pi)^{1/4}\over 7}(\Gamma M_{\rm Pl})^{1/2}~,
\end{equation}
with $\Gamma$ the decay width of the oscillating inflaton and the Higgs
fields into right-handed neutrinos~\cite{laz},
\begin{equation}
\label{g}
\Gamma={1\over 8\pi}\left({M_i\over M}\right)^2 m_{\rm infl}~,
\end{equation}
$m_{\rm infl}$ the inflaton mass, 
and $M_i$ the right handed neutrino mass eigenvalues.
Equations (\ref{Mdekappa}),~(\ref{rhn}),~(\ref{g}) lead to
 \begin{equation}
 T_{\rm RH}\sim {1\over 12}\left({60\over N_{\rm Q}}\right)^{1/4}
 \left({1\over { \cal N}}\right)^{1/4}    y_{\rm
 Q}^{1/2} M_i~.
 \end{equation}
The reheating temperature must satisfy the gravitino
constraint $T_{\rm RH}\leq 10^9$ GeV~\cite{gr}. This constraint
implies strong bounds on the $M_i$'s, which satisfy the inequality
$M_i < m_{\rm infl}/2$, where the inflaton mass is
\begin{equation}
\label{mgr}
m_{\rm infl}=\sqrt{2}\kappa M~.
\end{equation}
The strong bounds on $M_i$ lead to quite small corresponding
dimensionless couplings $\gamma_i$. The two heaviest
neutrino are expected to have masses of the order of $M_3\simeq
10^{15}$ GeV and $M_2\simeq 2.5 \times 10^{12}$ GeV
respectively~\cite{pati}. This implies that for $y_{\rm Q}$ of the
order of 1, $M_i$ in Eq.~(\ref{rhn}) cannot be identified with the
heaviest or the next to heaviest right handed neutrino, 
otherwise the reheating temperature would be higher than the upper
limit imposed by the gravitino bound. Thus, $M_i$ is $M_1$, with
$M_1\sim 6\times 10^9$ GeV~\cite{pati}.  (This value is in agreement
with the mass suggested in Ref.~\cite{laz} for the mass of the right handed 
neutrino into which the inflaton disintegrates.)

Therefore, using Eqs.~(\ref{rhn}),~(\ref{g}) and (\ref{mgr}), 
the gravitino constraint on the reheating temperature implies, 
\begin{equation}
\frac{\sqrt2}{14\pi^{1/4}}M_1\left[\frac{M_{\rm Pl}}{M(\kappa)}\right]^{1/2}
\sqrt\kappa\leq 10^9{\rm GeV}~.
\end{equation}
Using the fit of the function $M(\kappa)$ given in Eq.~(\ref{fitM}), it
is possible to evaluate numerically $\sqrt{\kappa/M(\kappa)}$ and find
an upper limit on the allowed values for the coupling of the superpotential, 

\begin{equation}
\kappa\lsim 8\times 10^{-3}~,
\end{equation}
in agreement with the upper bound found in Ref.~\cite{ss2004}.

Secondly, we proceed with a strongest constraint on $\kappa$, imposed
from the measurements on the CMB temperature anisotropies.  The cosmic
strings contribution to the quadrupole can be calculated from
\begin{equation}
{\cal A}_{\rm cs}\sim\left[
{\left({\delta T\over T}\right)^{\rm cs}_{\rm Q}
\over
\left({\delta T\over T}\right)^{\rm COBE}_{\rm Q}}\right]^2~,
\end{equation}
and it is shown in Fig.~\ref{contribCSplot}.

\begin{figure}[hhh]
\begin{center}
\includegraphics[scale=.6]{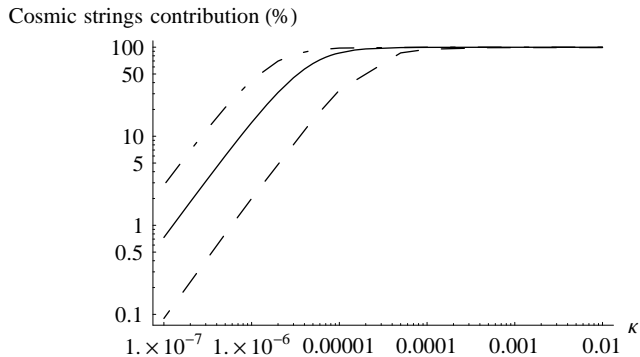}
\caption{Evolution of the cosmic strings contribution to the
quadrupole anisotropy as a function of the coupling  of the
superpotential, $\kappa$. The three curves correspond to
$\mathcal{N}=\mathbf{27}$ (curve with broken line), $\mathcal{N}=
\mathbf{126}$ (full line) and $\mathcal{N}=\mathbf{351}$
(curve with lines and dots).}\label{contribCSplot}
\end{center}
\end{figure}

As one can see, for small values of the coupling $\kappa$ the cosmic strings
contribution depends on the value of $\mathcal{N}$.  Thus, we should find the
value of $\mathcal{N}$ required for the allowed SSB patterns leading to
G$_{\rm SM}\times Z_2$.\\ Within the framework of our study, the SSB schemes
allowed from particle physics and cosmology are explicitely found in
Ref.~\cite{rjs} and we recall them in Appendix A. The parameter $\mathcal{N}$
is the dimensionality of the Higgs fields representation that generate the
SSB; these fields are coupled to the inflaton. As explained in
Ref.~\cite{rjs}, the inflationary era should take place after the last
formation of monopoles and/or domain walls since these objects are
incompatible with cosmology. Therefore, there are just very few choices for
the SSB stage where inflation can be placed.  The value of the parameter
$\mathcal{N}$ depends on the GUT gauge group and the SSB scheme where
inflation takes place. For SO(10), under the requirement that R-parity is
conserved down to low energies, $\mathcal{N}=\mathbf{126}$.  For
$\mathrm{E}_6$, the Higgs representations can be $\mathcal{N}=\mathbf{27}$ or
$\mathcal{N}=\mathbf{351}$. Our results depend slightly on the choice of
$\mathcal{N}$ and in what follows, we focus mainly on
$\mathcal{N}=\mathbf{126}$.

Comparing the results we obtained for the cosmic strings contribution
to the CMB fluctuations as a function of the superpotential coupling
$\kappa$, (see, Fig.~\ref{contribCSplot}) against the cosmic strings
contribution allowed from the measurements, we can constrain
$\kappa$. Already BOOMERanG~\cite{boom}, MAXIMA~\cite{maxi} and
DASI~\cite{dasi} experiments imposed~\cite{bouchet} an upper limit on
${\cal A}_{\rm cs}$, which is ${\cal A}_{\rm cs}\lsim 18\%$.
Clearly, the limit set by the Wilkinson Microwave Anisotropy Probe
(WMAP) measurements~\cite{wmap} should be stronger. A recent Bayesian
analysis in a three dimensional parameter space~\cite{pogosian} have
shown that a cosmic strings contribution to the primordial
fluctuations ${\cal A}_{\rm cs}$ higher than $9\%$ is excluded up to
$99\%$ confidence level.

 We state below the constraint we found for
the dimensionless parameter $\kappa$, as a function of $\mathcal{N}$,
\begin{equation}
\label{kappa_constr}
\kappa \lsim 7\times 10^{-7} \times \frac{126}{\mathcal{N}}~,
\end{equation}
which holds for $\mathcal{N} \in 
\{\mathbf{1}, \mathbf{16},\mathbf{27},\mathbf{126},\mathbf{351}\}$.
This constraint on $\kappa$, Eq.~(\ref{kappa_constr}), is in agreement
with the one found in Ref.~\cite{kl}.

Both, the CMB lower bound and the gravitino upper bound on
$\kappa$ are summarised in Fig.~\ref{contraintekappa}.

\begin{figure}[hhh]
\begin{center}
\includegraphics[scale=.6]{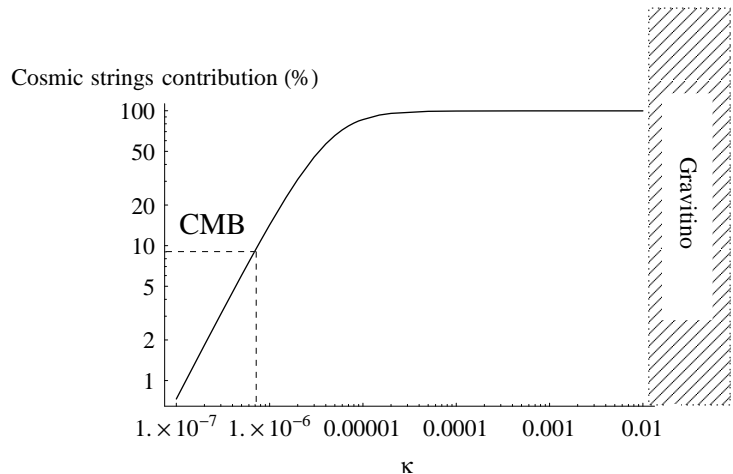}
\caption{Constraints on the single parameter $\kappa$ of the
model. The gravitino constraint implies     
$\kappa\leq 8\times 10^{-3}$. The allowed cosmic strings 
contribution to the CMB angular power spectrum implies
$\kappa \lsim 7\times 10^{-7}$, for $\mathcal{N}=\mathbf{126}$.}
\label{contraintekappa}
\end{center}
\end{figure}

The implications of this  constraint on the single free
parameter of the F-term inflationary model are important. Hybrid
F-term inflation was known to have an appealing feature as compared to
the most elegant inflationary scenario within nonsupersymmetric
theories, i.e. chaotic inflation.  Namely, it was believed that there
was no need of fine tuning which will set a very small value for the
coupling  of the superpotential.  This nice feature has been
disappeared once we compare theoretical predictions against
cosmological data. As we show below, one possible way out is to employ
the curvaton mechanism.  Another implication of this
constraint, quite important for cosmology, is the fact that this
constraint on $\kappa$ can be converted into a constraint on the mass
parameter $M$. We would like to remind to the reader that this parameter 
controls the mass of the strings formed, as well as the inflationary scale.
Using the CMB limit on the cosmic strings contribution, we can also 
obtain that, up to 99\% of confidence level,
\begin{equation}
M \lsim 2\times 10^{15}~{\rm GeV}~.
\end{equation}
 As we show below, this constraint is robust since it holds even when
the curvaton mechanism is involved, and for all  values of $\mathcal{N}$.

In the curvaton scenario, the curvaton field is responsible for the
generation of the primordial fluctuations. The curvaton is a scalar
field, that is subdominant during the inflationary era as well as at
the beginning of the radiation dominated era which follows the
inflationary phase. The effective curvaton mass is assumed to be much
smaller that the Hubble parameter during the inflationary phase.
Within the framework of supersymmetry, which is the context of our
study, one expects the existence of scalar fields. Thus, it is
reasonable to expect that one of them could indeed play the r\^ole of
the curvaton field.  In addition, in the class of models we are
considering, one may expect that it may exist a natural candidate for
the curvaton field. As we have shown in Ref.~\cite{rjs}, in many
acceptable SSB schemes apart from the formation of topological cosmic
strings we have the formation of embedded strings, which are
topologically and dynamically unstable.  If the decay product of
embedded strings can give a scalar field before the onset of
inflation, then such a scalar field could play the r\^ole of the
curvaton field.

Assuming the existence of a curvaton field, there is an additional
contribution to the temperature anisotropies. Thus,
\begin{equation}\label{equacurv}
\left[\left(\frac{\delta T}{T}\right)_{\rm tot}\right]^2=
\left[\left(\frac{\delta T}{T}\right)_{\rm infl}\right]^2+
\left[\left(\frac{\delta T}{T}\right)_{\rm cs}\right]^2
+\left[\left(\frac{\delta T}{T}\right)_{\rm curv}\right]^2~.
\end{equation}
The inflaton contribution has a scalar and a tensor part, however
since the tensor part in the supersymmetric hybrid inflation is
always much smaller than the scalar one, we neglect it.

Since the primordial curvaton fluctuation is converted to purely
adiabatic density fluctuations, the curvaton contribution in terms of
the metric perturbation reads
\begin{equation}
\left(\frac{\delta T}{T}\right)_{\rm curv} \equiv \frac{\Psi_{\rm curv}}{3}~,
\end{equation}
and from  Eq.~(\ref{Psi_curv}) we have
\begin{equation}
\left(\frac{\delta T}{T}\right)_{\rm curv}= -\frac{4}{27}\frac{\delta
{\cal\psi}_{\rm init}}{{\cal\psi}_{\rm init}}~.
\label{above}
\end{equation}
For $V\simeq \kappa^2 M^4$,
Eqs.~(\ref{qfc}),~(\ref{exprate}),~(\ref{Mdekappa}) and (\ref{above})
imply
\begin{equation}\label{curvterm}
\left[\left({\delta T\over T}\right)_{\rm curv}\right]^2 =
y_Q^{-4}\,\left({\kappa^2 \mathcal{N} N_{\rm Q}\over 32\pi^2} \right)^2\, 
\left[ \left(\frac{16}{81\pi\sqrt3}\right)\,
\kappa\, \left(\frac{M_{\rm Pl}}{{\cal\psi}_{\rm init}}\right)\right]^2~,
\end{equation}

\begin{figure}[hhh]
\begin{center}
\includegraphics[scale=.6]{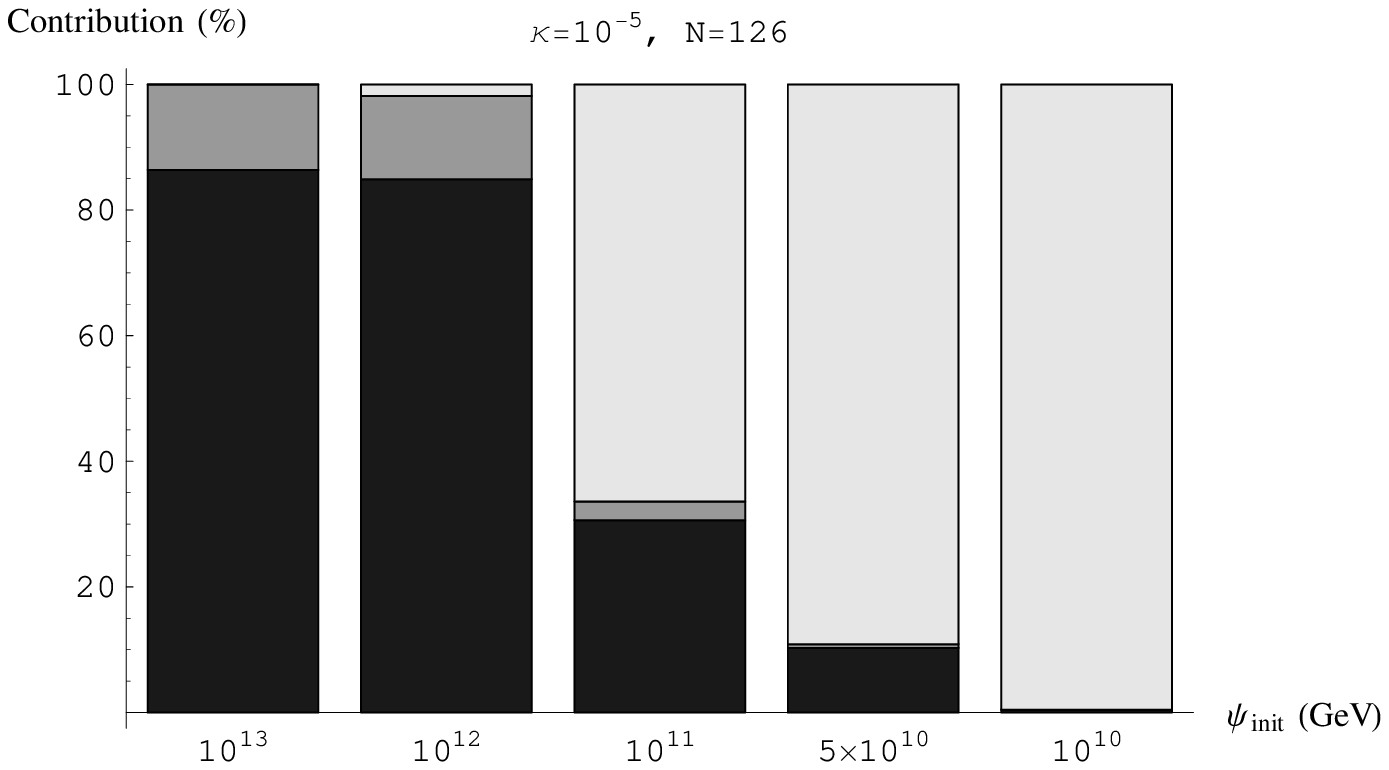}
\includegraphics[scale=.55]{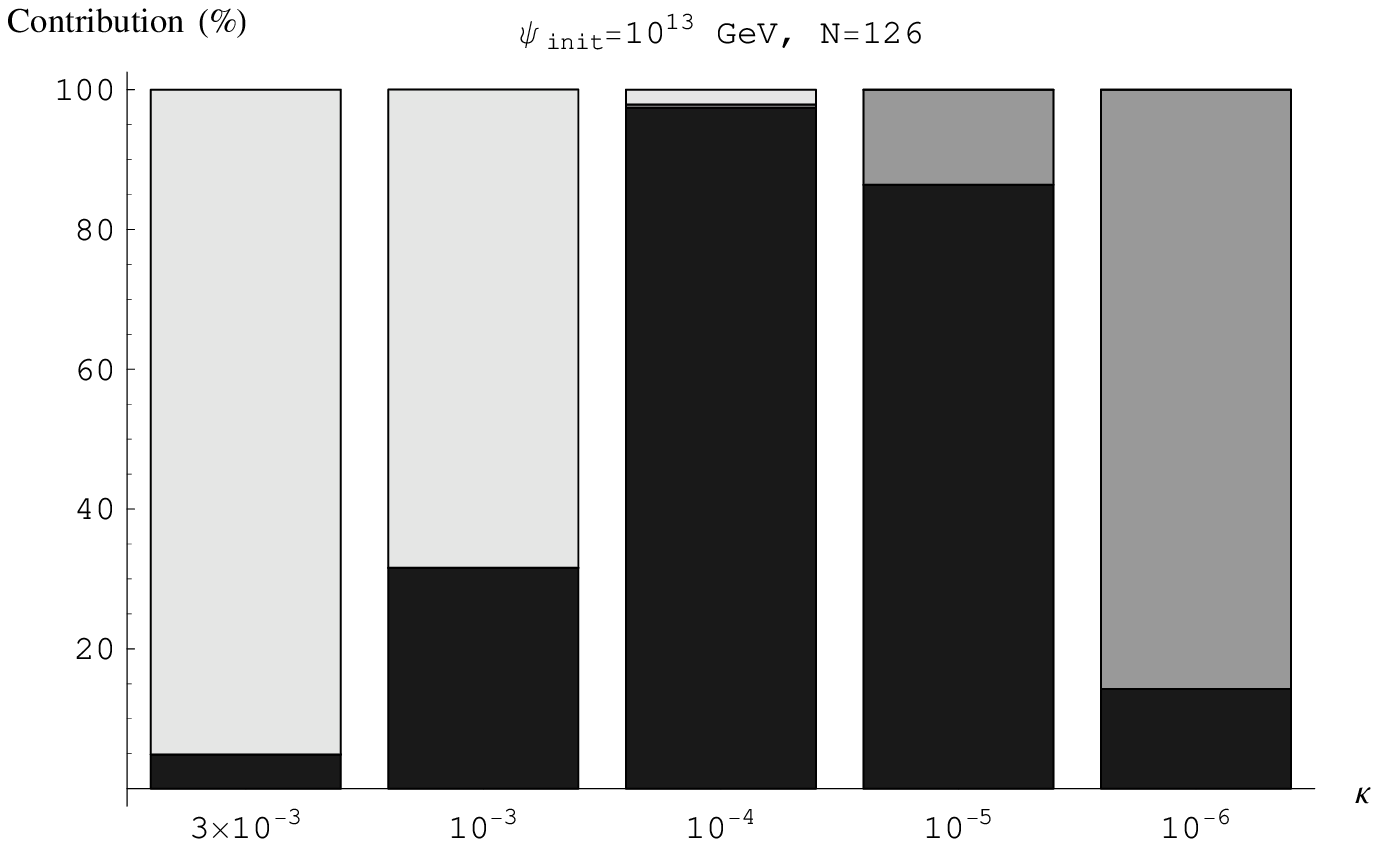}
\caption{The cosmic strings (dark gray), curvaton (light gray) and
inflaton (gray) contributions to the CMB temperature anisotropies as a
function of the the initial value of the curvaton field
${\cal\psi}_{\rm init}$, and the superpotential coupling 
$\kappa$, for ${\cal N}=\mathbf{126}$.}
\label{contribcurv}
\end{center}
\end{figure}

We solve Eq.~(\ref{equacurv}) using Eq.~(\ref{curvterm}) and we obtain
the different contributions of the three sources of anisotropies:
inflaton, cosmic strings and curvaton.  Their respective contributions
as a function of $\kappa$, or ${\cal\psi}_{\rm init}$, are shown in
Fig.~\ref{contribcurv}.  for a fixed value of
$\kappa$, the cosmic strings contribution decreases as
${\cal\psi}_{\rm init}$ decreases, while the curvaton contribution
becomes dominant.  It is thus possible to use the WMAP constraint to
limit ${\cal\psi}_{\rm init}$. The upper bound on ${\cal\psi}_{\rm
init}$  depends on the coupling  $\kappa$. In addition, one has to impose
the gravitino constraint.  Therefore, a coupling
 bigger than $10^{-2}$ is excluded. More precisely,
\begin{equation}
{\cal\psi}_{\rm init} \lsim 5\times 10^{13}\,\left( 
\frac{\kappa}{10^{-2}}\right){\rm GeV}~.
\end{equation}
The above constraint is valid only if $\kappa$ is in the range
$[10^{-6},~1]$, while for smaller values of $\kappa$, the
cosmic strings contribution is smaller than the WMAP limit for any
value of the curvaton field (see, Fig.~\ref{contribcurv}).

\subsection{F-term inflation and supergravity corrections}

We proceed with the supergravity corrections. 
The scalar potential in supergravity has the general form~\cite{nilles}
\begin{equation}\label{potenSUGRA}
V=\frac{e^G}{M_{\rm Pl}^4}\left[G_i(G^{-1})^i_jG^j-3 \right] \quad
\mathrm{with}\quad G=\frac{K}{M_{\rm Pl}^2}+\ln\frac{|W|^2}{M_{\rm
Pl}^6}~,
\end{equation}
where the K$\ddot{\rm a}$hler potential $K(\phi, \phi^*)$ is
a real function of the complex scalar fields $\phi_i$,
and their Hermitian conjugates ${\phi_i}^*$. 
The $\phi_i$ are scalar components of the chiral superfields $\Phi_i$.
We adopted the following notations
\begin{equation}
G^i\equiv \frac{\partial G}{\partial \phi_i}~, 
\quad G_j\equiv \frac{\partial G}{\partial {\phi^j}^*}~.
\end{equation}

Assuming that the D-term is flat along the inflationary trajectory,
one can ignore it during F-term inflation. Thus, we only consider the
F-term in the above equation. If we expand $K=\sum_i
|\phi_i|^2+\cdots$, the corrections we get to the lowest order
inflationary potential lead to the scalar potential
\begin{equation}
  V=\sum_i\left|{\partial W\over
\partial\phi_i}\right|^2\left(1+{1\over M_{\rm
Pl}^2}\sum_i|\phi_i|^2+\cdots\right)+{\rm other
~terms}~.
\end{equation}
As it was shown in
Ref.~\cite{sugra}, assuming the minimal form for the K$\ddot{\rm a}$hler 
potential $K$ and minimising the scalar potential $V$ with respect to 
${\phi_-}$ and $\phi_+$ for $|S|>M$, we obtain that the
SUGRA correction to the scalar potential is
\begin{equation}
V_{\rm SUGRA}=\kappa^2 M^4\left[{1\over 8} {|S|^4\over M_{\rm
Pl}^4}+\cdots\right]~.
\end{equation}
The effective scalar potential becomes
$$ V_{\rm eff}=\kappa^2 M^4
\biggl\{1+\frac{\kappa^2\cal{N}}{32\pi^2}\bigl[2\ln\frac{|S|^2\kappa^2}
{\Lambda^2}
+(z+1)^2\ln(1+z^{-1})+(z-1)^2\ln(1-z^{-1})\bigr]\nonumber
$$
\begin{equation}
~~~~~~~~~~+ {1\over 8} {|S|^4\over M_{\rm Pl}^4}\biggl\}~.
\end{equation}

In the literature (e.g., see Ref.~\cite{DvaShaScha}), some authors
consider only two terms for the scalar potential. Namely, they
consider only the first term of the radiative corrections and the term
coming from the SUGRA correction. However, this assumption holds only
if $x_{\rm Q}\gg 1$. In our study we have found that the order of 
magnitude for $x_{\rm Q}$ is ${\cal O}(1)$ except for high values of 
the superpotential coupling $\kappa$
(namely $\kappa \geq 10^{-2}$), but those are forbidden by the
gravitino constraint. In conclusion, we find that all  terms coming from 
the radiative corrections have to be taken into account.

Since $|S_{\rm Q}|/M_{\rm Pl}\lsim 10^{-3}$, the potential and its first 
derivative is not significantly affected by the SUGRA corrections. Thus, 
including these corrections we do not expect to find
any difference in the $\delta T/T$. This is indeed what one expects,
since the value of the inflaton field, in the absence of SUGRA
corrections, stays far below the Planck mass.

In conclusion, one can indeed neglect the supergravity corrections in
the framework of F-term inflation.

\subsection{D-term inflation in global supersymmetry}

In what follows, we derive in the framework of global SUSY the cosmic
strings contribution to the temperature anisotropies of the CMB.  We
use $V\simeq V_0=g^2\xi^2/2$, while we employ Eq.~(\ref{VexactD}) to
derive the exact expression for $V'\equiv \partial V(|S|)/\partial
|S|$, which reads
\begin{equation}
V'(|S|)=\frac{2b}{|S|}z\,f(z)~,
\end{equation}
where $f(z)$ is the same as in the F-term, Eq.~(\ref{fdez}), and
\begin{equation}
b=\frac{g^4\xi^2}{16\pi^2}.
\end{equation}
The next step is to express the number of e-foldings $N_{\rm Q}$.
Employing the expressions written above  for $V$ and $V'$ and using
Eq.~(\ref{defnefold}), we obtain 
\begin{equation}
N_{\rm Q}=\frac{2\pi^2}{\lambda^2}
\frac{\xi}{M_{\rm Pl}^2} \int_{z_{\rm end}}^{z_Q} \frac{{\rm d}z}{zf(z)}~,
\end{equation}
with 
\begin{equation}
z_{\rm end}={\lambda^2|S_{\rm end}|^2\over g^2\xi}~.
\end{equation}
One has to evaluate the value of the
inflaton field at the end of inflation.  The inflationary era ends
when one of the two following conditions is reached:
\begin{itemize}
\item In the global minimum of the potential, $\langle\phi_+\rangle$
or $\langle \phi_+ \rangle$ is not zero anymore.  This means that one of
the two fields acquires a positive effective mass squared.  Let
us call $z_{\rm SB}$ the value of $z$ for which this condition is
realized.
\item The slow roll conditions cease to be satisfied.  Within
supersymmetric hybrid inflation, the slow roll parameter $\epsilon$ is
generically very small, and the slow roll condition becomes just
$\eta\sim 1$.  We denote by $z_{\rm SR}$ the value of $z$ for which
the slow roll condition ceases to be satisfied.\\
\end{itemize}
Thus,
\begin{equation}\label{endinfl}
z_{\rm end}=\max(z_{\rm SB},z_{\rm SR})~.
\end{equation}
Let us calculate the values of $z_{\rm SB}$ and $z_{\rm SR}$.
Firstly, $z_{\rm SB}$ can be read from the quadratic part
(for the fields $\phi_+$ and $\phi_-$) of the potential as it is given
by Eq.~(\ref{VtotD}). Namely, from the Lagrangian
\begin{equation}
\mathcal{L}=\lambda^2 |S|^2(|\phi_+|^2+|\phi_-|^2)
+g^2\xi(|\phi_+|^2-|\phi_-|^2)~,
\end{equation}
we get $z_{\rm SB}=1$.\\
Secondly, the end of the slow roll phase of inflation is reached when
the slow roll parameter $\eta$ becomes of the order of unity, 
\begin{equation}\label{equaeta}
\eta\equiv M_{\rm Pl}^2\left[\frac{V''(S)}{V(S)}\right]\sim 1~.
\end{equation}
For our model, 
\begin{equation}
\eta=\frac{\lambda^2}{4\pi^2}\left(\frac{M_{\rm Pl}^2}{\xi}\right)g(z)~,
\end{equation}
where
\begin{equation}\label{gdez}
g(z)=(3z+1)\ln(1+z^{-1})+(3z-1)\ln(1-z^{-1})~.
\end{equation}

Equation (\ref{equaeta}) can be solved provided $ \lambda \lsim 4
\times 10^{-3}$. This is a reasonable condition since
higher values of $\lambda$ would be forbidden by the gravitino
constraint (see discussion in III.A). For $\lambda \lsim 4
\times 10^{-3}$, $\xi/\lambda^2 \gsim 10^{36}$ and as one can
see from the shape of the function $g(z)$,
the two solutions of Eq.~(\ref{equaeta}) are very close to $1$. This 
implies $z_{\rm SR}\sim 1$, and the number of e-foldings reads
\begin{equation}\label{NQ-D}
N_{\rm Q}=
\frac{2\pi^2}{\lambda^2}\, \frac{\xi}{M_{\rm Pl}^2}\, y_{\rm Q}^2~,
\end{equation}
where $y_{\rm Q}$ is given by Eq.~(\ref{yQ}) with $x_{\rm
Q}=\lambda|S_{\rm Q}|/(g\sqrt{\xi})$.

To measure the weight of the cosmic strings contribution,
we suppose three sources for the quadrupole anisotropy of the CMB, namely
\begin{equation}
\left[\left(\frac{\delta T}{T}\right)_{\rm tot}\right]^2=
\left[\left(\frac{\delta T}{T}\right)_{\rm scal}\right]^2
+\left[\left(\frac{\delta T}{T}\right)_{\rm tens}\right]^2+
\left[\left(\frac{\delta T}{T}\right)_{\rm cs}\right]^2~,
\end{equation}
and we calculate each term using
Eqs.~(\ref{contribInflScal}),~(\ref{contribInflTens}),~(\ref{contribCS})
respectively. In this last equation, the VEV of the Higgs field
responsible of the cosmic strings formation is $\sqrt\xi$.\\ Thus,
eliminating $\xi$ with Eq.~(\ref{NQ-D}), we obtain
\begin{eqnarray}\label{eqnumD}
\left({\delta T\over T}\right)_{\rm Q- tot} &\sim&
\Big\{y_{\rm Q}^{-4}\left({\lambda^2 \,
N_{\rm Q}\over 16\pi^2}\right)^2
\Big[\frac{16N_{\rm Q}}{45} x_{\rm Q}^{-2}y_{\rm Q}^{-2}f^{-2}(x_Q^2)
\nonumber\\
&& +\left(\frac{0.77 g}{\sqrt{2}\pi}\right)^2 + 324\Big]\Big\}^{1/2}~,
\end{eqnarray}
where $\left(\delta T/ T\right)_{\rm Q}^{\rm tot}$ is normalised to
the COBE data, i.e., $\left(\delta T/ T\right)_{\rm Q}^{\rm COBE} \sim
6.3\times 10^{-6}$.  For given values of $\lambda$ and $N_{\rm Q}$ we
solve numerically Eq.~(\ref{eqnumD}) to get $x_{\rm Q}$.

Regarding D-term inflation we find that, as in the F-term case,
the tensor contribution to the ${\delta T/ T}$ is negligible.  For
$\lambda \lsim 4\times 10^{-3}$, one can compute the
mass scale $M_{\rm D}=\sqrt\xi$ as a function of $\lambda$.  We
find that $\xi$, which can be seen as a {\sl mass parameter}, has a
very similar shape as the one found for $M$ in the case of F-term
inflation (see, Fig.~\ref{inflscaleF}).  Moreover, we obtain that the
contribution of the cosmic strings to the quadrupole anisotropy is
very similar to the F-term case.  In addition, to be consistent with
the CMB data, $\lambda$ in D-term inflation should obey a similar
limit to the one found for $\kappa$ in F-term inflation, namely $\lambda 
\lsim 3\times 10^{-5}$. These results
are summarised in Fig.~\ref{dterm}.

\begin{figure}[hhh]
\begin{center}
\includegraphics[scale=.5]{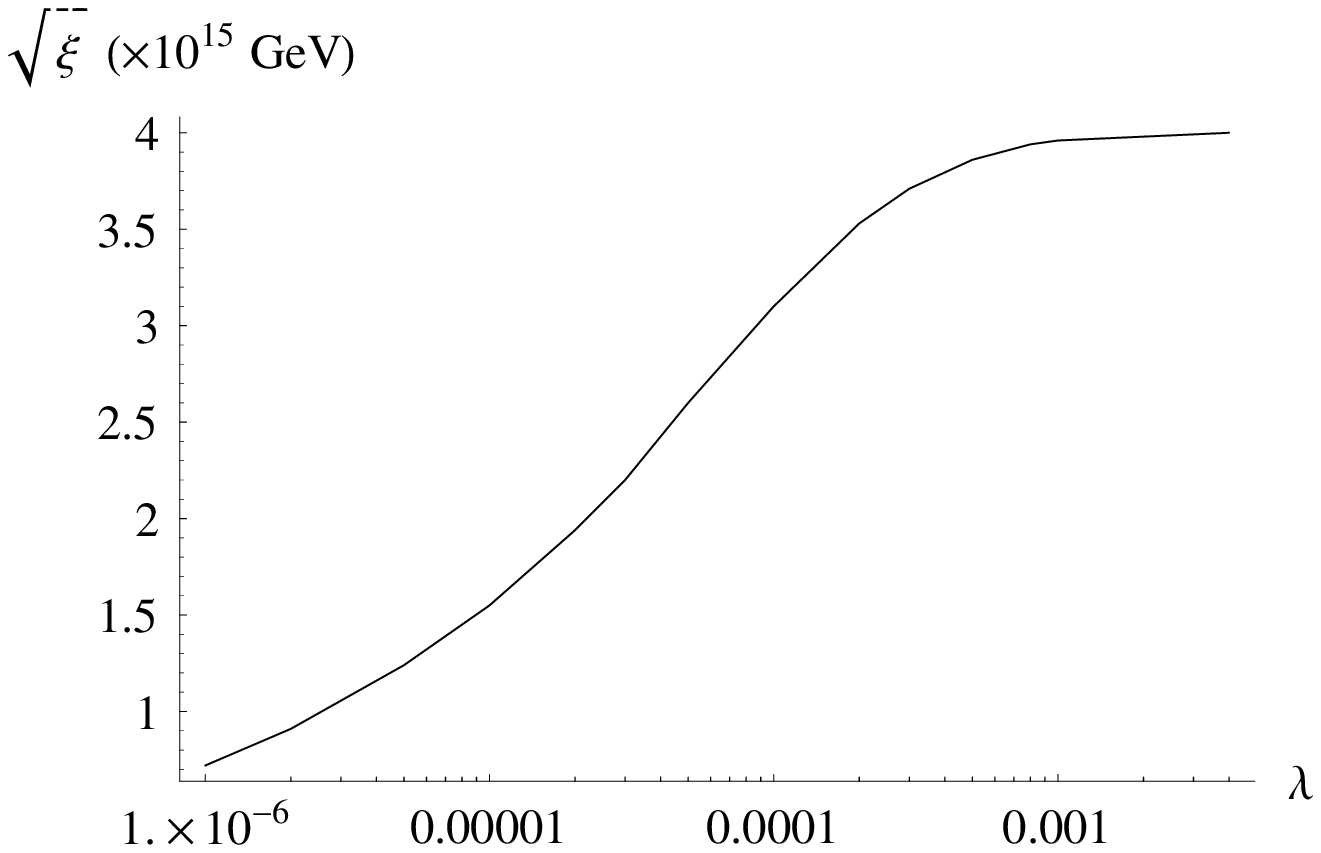}
\includegraphics[scale=.5]{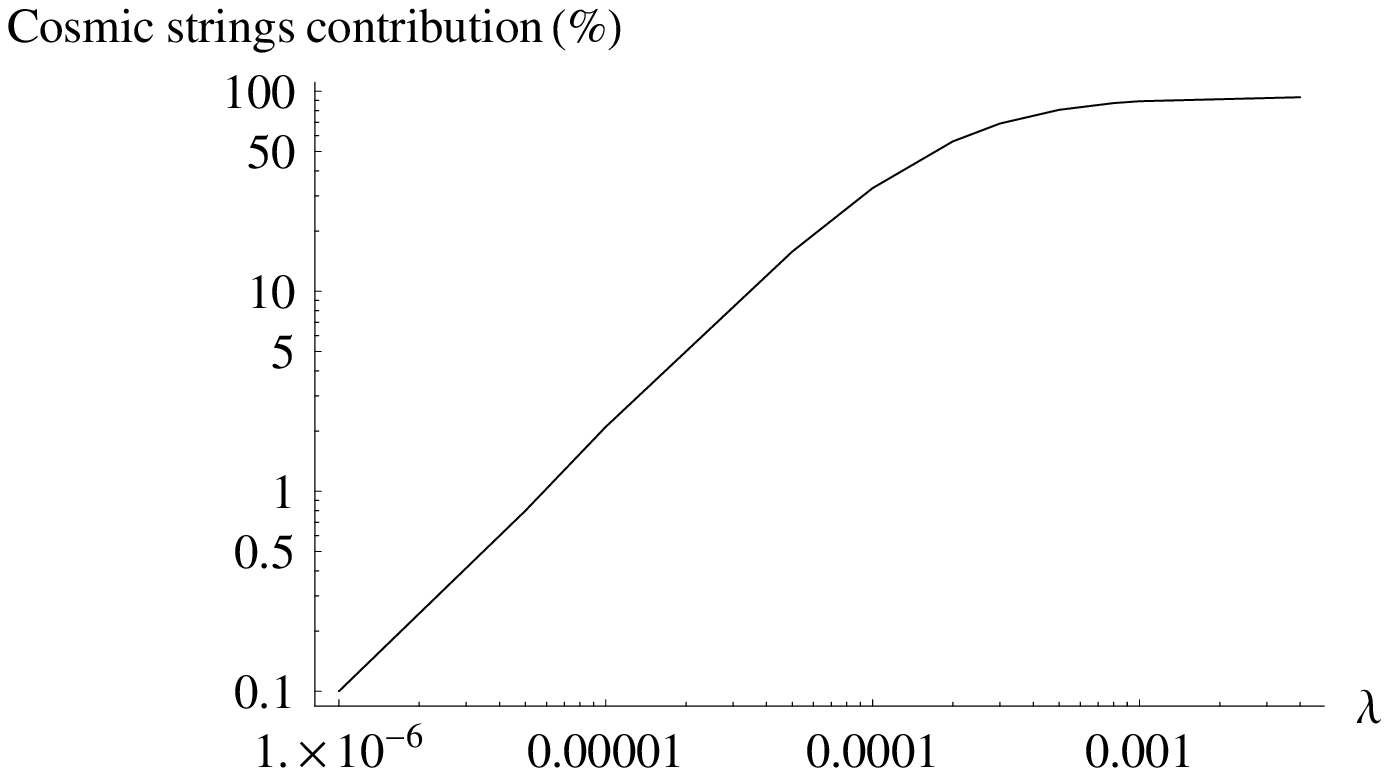}
\caption{On the left, evolution of the mass scale $\sqrt\xi$ as a function 
of the coupling  $\lambda$. On the right, evolution of the cosmic
strings contribution to the quadrupole anisotropy as a function of the
coupling  of the superpotential, $\lambda$. These plots are derived in the 
framework of SUSY.}
\label{dterm}
\end{center}
\end{figure}

At this point we would like to bring to the attention of the reader
that our findings regarding D-term inflation in SUSY, disagree with
some results stated in previous studies\footnote{More precisely, we disagree
with the statement that {\sl in the context of D-term inflation cosmic strings
contribute to the $C_\ell$'s up to the level of $75\%$.}} (see eg.,
Ref.~\cite{RJ1}). The reason is that in our analysis, we
take into account {\sl all terms of the radiative corrections} and, as we
have already shown in our discussion on F-term inflation, these terms
have an essential impact on the determination of the cosmic strings
contribution to the CMB.

Knowing $x_{\rm Q}$, one can compute from Eq.~(\ref{xdeS}) the order
of magnitude of the corresponding inflaton field $S_{\rm Q}$.  Even
though Eq.~(\ref{eqnumD}) is very similar to the one we had in the
case of F-term hybrid inflation, i.e., Eq.~(\ref{eqnumF}), the
conclusions for the value of the inflaton field as compared to the
Planck mass are quite different.  This is due to the dependence of
$x_{\rm Q}$ on the coupling $\lambda$, which makes the inflaton field
$S_{\rm Q}$ to be of the order of the Planck mass or higher.

The analysis we presented above implies that global supersymmetry is
insufficient  to study D-term inflation. It is clear that only 
in the context of supergravity one can treat correctly the issue 
of D-term inflation and its cosmological implications. 

\subsection{D-term inflation in supergravity}

The case of D-term inflation has to be addressed within the framework
of supergravity, since as we have shown in the previous subsection,
the fields are not negligible as compared to the Planck mass.

Inflation is still derived from the superpotential
\begin{equation}
W^{\rm D}_{\rm infl}=\lambda S \Phi_+\Phi_-~.
\end{equation}
The F-term part of the scalar potential is given by 
Eq.~(\ref{potenSUGRA}). Considering also the D-term, the total scalar 
potential reads
\begin{equation}\label{DpotenSUGRA}
V=\frac{e^G}{M_{\rm Pl}^4}\left[G_i(G^{-1})^i_jG^j-3 \right] + \frac{1}{2} 
[{\rm Re} f(\Phi_i)]^{-1} \sum_a g_a^2 D_a^2~,
\end{equation}
where $f(\Phi_i)$ is the gauge kinetic function and 
\begin{equation}
G=\frac{K}{M_{\rm Pl}^2}+\ln\frac{|W|^2}{M_{\rm Pl}^6}~.
\end{equation}
The K$\ddot{\rm a}$hler potential $K(\phi_i, \phi_i^*)$ is a real
function of the complex scalar fields $\phi_i$, and their Hermitian
conjugates ${\phi_i}^*$ where the $\phi_i$ are scalar components of the chiral
superfields $\Phi_i$.  Upper (lower) indices $(i,j)$ denote
derivatives with respect to $\phi_i$ (${\phi^i}^*$).  For the D-term
part, $\xi_a$ is the Fayet-Iliopoulos term, $g_a$ the coupling
 of the $U(1)^a$ symmetry, which is generated by ${T_a}$ and
under which the chiral superfields $S$, $\Phi_+$ and $\Phi_-$ have
charges $0$, $+1$ and $-1$ respectively. Finally,
\begin{equation}
D_a=\phi_i {(T_a)^i}_j K^j+\xi_a~.
\end{equation}
In what follows, we assume\footnote{This is the simplest supergravity
model and in general the K$\ddot{\rm a}$hler potential can be a more
complicated function of the superfields. This is generally the case
for supergravity theories derived from superstring theories.} the
minimal structure for $f(\Phi_i)$ (i.e., $f(\Phi_i)$=1) and take the 
minimal K$\ddot{\rm a}$hler potential given by
\begin{equation}\label{miniK}
K=|\phi_-|^2+|\phi_+|^2+|S|^2~.
\end{equation}
Therefore, as it was found in \cite{DSUGRA,halyo}, the scalar potential reads
\begin{eqnarray}\label{DpotenSUGRAtot}
V^{\rm D}_{\rm SUGRA} & =&
\lambda^2\exp\left({\frac{|\phi_-|^2+|\phi_+|^2+|S|^2}{M^2_{\rm
Pl}}}\right)
\biggl[|\phi_+\phi_-|^2\left(1+\frac{|S|^4}{M^4_{\rm
Pl}}\right)
\nonumber\\
&+&
|\phi_+S|^2 \left(1+\frac{|\phi_-|^4}{M^4_{\rm
Pl}}\right)+|\phi_-S|^2 \left(1+\frac{|\phi_+|^4}{M^4_{\rm Pl}}\right)
+3\frac{|\phi_-\phi_+S|^2}{M^2_{\rm Pl}}\biggl]\nonumber \\ &+&
\frac{g^2}{2}\left(|\phi_+|^2-|\phi_-|^2+\xi\right)^2~.
\end{eqnarray}

As in the case of global supersymmetry, for $S>S_{\rm c}$ the potential is
minimised for $|\phi_+|=|\phi_-|=0$ and therefore, at the tree level,
the potential in the inflationary valley is constant,
$V_0=g^2\xi^2/2$. However, despite what is sometines claimed, this does
not mean that the effective inflationary potential is identical to
the one found in the case of global supersymmetry. One has to compute
the radiative corrections, and must first  calculate the effective masses of
the components of the superfields $\Phi_+$ and $\Phi_-$. Extracting the 
quadratic terms from the potential, given by Eq.~(\ref{DpotenSUGRAtot}), 
 the scalar components $\phi_+$ and $\phi_-$ get squared masses
\begin{equation}\label{massbos}
m^2_{\pm}=\lambda^2|S|^2 \exp\left({\frac{|S|^2}{M^2_{\rm
Pl}}}\right)\pm g^2 \xi~.
\end{equation}
To calculate the masses of the fermionic components $\Psi_+$ and
$\Psi_-$, one has to use the Lagrangian of the fermionic sector,
\begin{equation}
\mathcal{L}_{\rm fermion}=\frac{1}{2}e^{G/2}\left[-G^{ij}-G^{i}G^{j}+
G^{ij}_k(G^{-1})^k_lG^l\right]\bar{\Psi}_{iL}\Psi_{jR} + {\rm h.c.}~,
\end{equation}
where $\Psi_{iL}$ and $\Psi_{jR}$ are left and right components of the
Majorana spinors $\Psi_+$ and $\Psi_-$.  Using Eq.~(\ref{miniK}), one
gets a Dirac fermion with squared mass
\begin{equation}\label{massferm}
m_{\rm f}^2=\lambda^2|S|^2 \exp\left({\frac{|S|^2}{M^2_{\rm Pl}}}\right)~.
\end{equation}
We calculate the radiative corrections $[\Delta V(|S|)]_{1-{\rm loop}}$ 
using 
the Coleman-Weinberg expression, Eq.~(\ref{cw}).
Thus, we compute the full effective scalar potential during D-term 
inflation, within the context of SUGRA
\begin{eqnarray}
\label{VexactDsugra}
V^{\rm D-SUGRA}_{\rm eff}&=& V_0 + [\Delta V(|S|)]_{1-{\rm loop}}\nonumber\\
&=&
\frac{g^2\xi^2}{2}\left\{1+\frac{g^2}{16\pi^2}
\biggl[2\ln\frac{\lambda^2|S|^2} {\Lambda^2} \exp
\biggl({\frac{|S|^2}{M^2_{\rm Pl}}}\right)\nonumber\\
&&~~~+
(z+1)^2\ln(1+z^{-1})+(z-1)^2\ln(1-z^{-1})\biggl]\biggl\}
\end{eqnarray}
with
\begin{equation}\label{xdeSsugra}
z=\frac{\lambda^2 |S|^2 }{g^2\xi} \exp\left({\frac{|S|^2}{M^2_{\rm
Pl}}}\right)\equiv x^2 ~.
\end{equation}

The above potential is a generalisation of the
SUSY D-term potential given by Eq.~(\ref{VexactD}).  The end of
inflation is achieved when one of the two conditions is satisfied: (i)
the symmetry is spontaneously broken, or (ii) the slow roll condition
is violated.  By definition of the variable $z$, the symmetry breaking
condition is still
\begin{equation}
z_{\rm SB}= 1~.
\end{equation}
For the slow roll condition, one has to compute the slow roll parameter
$\eta$ and then to solve $\eta\sim 1$. With the effective potential
given by Eq.~(\ref{VexactDsugra}), one can compute its first derivative 
\begin{equation}\label{derivativeSUGRA}
V'(|S|) = \frac{2b}{|S|}\, zf(z)\left(1+\frac{|S|^2}{M_{\rm Pl}^2}
\right)~.
\end{equation}
Assuming $V\simeq V_0$ and using the exact expression for $V'$, 
the $\eta$ parameter reads
\begin{equation}
\eta(z) = \left(\frac{g^2}{16\pi^2}\frac{M_{\rm
Pl}^2}{|S|^2}\right)\,z\left[g(z)+\frac{|S|^2}{M_{\rm Pl}^2}
h_3(z)+\frac{|S|^4}{M_{\rm Pl}^4} h_4(z) \right]~,
\end{equation}
where $g(z)$ has been introduced for D-term inflation in Eq.~(\ref{gdez}), 
and $h_3(z), h_4(z)$ are given by
\begin{eqnarray}
h_3(z)&=&(9z+5)\ln (1+z^{-1}) +(9z-5)\ln (1-z^{-1})~,\\
h_4(z)&=&(4z+2)\ln (1+z^{-1}) +(4z-2)\ln (1-z^{-1})~.
\end{eqnarray}
Supergravity corrections become essential for $S$ at least one order of 
magnitude bigger that $M_{\rm Pl}$, and under this condition the solutions of
$\eta(z)=1$ are $z_{\rm SR}^{(1)}= 1^+$ and $z_{\rm SR}^{(2)}= 
1^-$. Thus,  $z$ at the end of inflation,
 as defined by Eq.~(\ref{endinfl}), is $z_{\rm end}=1$.

The more complicated expression for $z(|S|)$ makes the calculation of
the different contributions to the CMB power spectrum trickier. The
cosmic strings contribution remains the same as the one computed in
the case of SUSY D-term inflation, since it depends only on the scale
$\sqrt\xi$. Concerning the inflaton contribution, we can use the 
dominant term of the potential $V\simeq V_0$ and the exact expression 
for the first derivative, Eq.~(\ref{derivativeSUGRA}), to write the number 
of e-foldings
\begin{equation}
N_{\rm Q}={2\pi^2\over g^2}\,{\tilde y}_{\rm Q}(x_{\rm Q},\lambda,\xi,g)~,
\end{equation}
where 
\begin{equation}\label{y'Q}
{\tilde y}_{\rm Q}(x_{\rm Q},\lambda,\xi,g)=\int_1^{x_{\rm Q}^2} 
\frac{{\rm W}(z(g^2 \xi)/(\lambda^2 M_{\rm Pl}^2))}
{z^2 f(z)[1+{\rm W}(z(g^2 \xi)/(\lambda^2 M_{\rm Pl}^2))]^2}\, {\rm d}z~.
\end{equation}

We note that in the above definition of ${\tilde y}_{\rm Q}$, the function
${\rm W}(x)$ is the ``W-Lambert function'', i.e., the inverse of the
function $F(x)=xe^x$.  Setting $c\equiv (g^2 \xi)/(\lambda^2 M_{\rm
Pl}^2)$, the number of e-foldings $N_{\rm Q}$ becomes a function of 
only $c$ and $x_{\rm Q}$, once $g$ is fixed, 
and it is shown in Fig.~\ref{surface}.  Imposing $N_{\rm Q}=60$
allows us to find a numerical relation between $c$ and $x_{\rm Q}$, as
illustrated in Fig.~\ref{surface}.

\begin{figure}[hhh]
\begin{center}
\includegraphics[scale=.35]{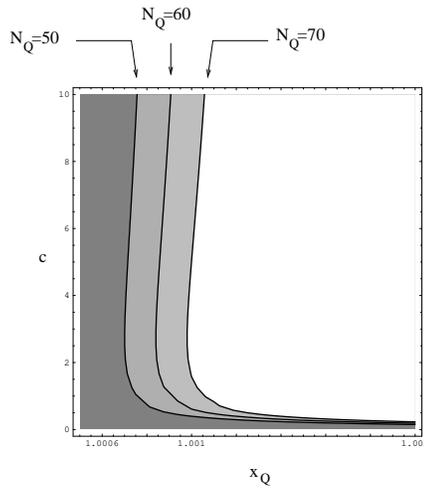}
\caption{Iso-contours of $N_{\rm Q}$ 
for $N_{\rm Q}=50,60,70$ in the plan $(x_\mathrm{Q},c\equiv 
(g^2 \xi)/(\lambda^2 M_{\rm Pl}^2))$. This graph is obtained for 
$g=10^{-2}$.}\label{surface}
\end{center}
\end{figure}

Using this relation between $c$ and $x_{\rm Q}$, it is possible to
construct a function $x_{\rm Q}(\xi)$ and then express the three
contributions (scalar, tensor and cosmic strings) to the CMB
temperature anisotropies only as a function of the Fayet-Iliopoulos
term $\xi$. Thus, we obtain that the total quadrupole temperature
anisotropy reads
\begin{eqnarray}\label{eqnumDsugra}
\left({\delta T\over T}\right)_{\rm Q}^{\rm tot} &\sim&
\frac{\xi}{M_{\rm Pl}^2}\Big\{ \frac{\pi^2}{90g^2}x^{-4}_{\rm
Q}f^{-2}(x^2_{\rm Q}) \frac{{\rm W}(x^2_{\rm Q}(g^2 \xi)(\lambda^2 M_{\rm
Pl}^2))} {\left[1+{\rm W}(x^2_{\rm Q}(g^2 \xi)(\lambda^2 M_{\rm
Pl}^2))\right]^2}\nonumber\\
&&
 +\left(\frac{0.77
g}{8\sqrt{2}\pi}\right)^2 +
\left(\frac{9}{4}\right)^2\Big\}^{1/2}~,
\end{eqnarray}
where the only unknown is $\xi$, for given values of $g$ and
$\lambda$.  We normalise the quadrupole anisotropy to COBE and we
calculate numerically $\xi$, and thus $x_{\rm Q}$ using the function
$x_{\rm Q}(\xi)$, as well as the various
contributions as a function of the couplings $\lambda$ and $g$.

Our results are listed below. Firstly, as previously, it is
straightforward to see that the tensor contribution, even for $g=1$,
is completely negligible. Second, the inflaton field $S_{\rm Q}$ is
of the order of $10 M_{\rm Pl}$ for the studied parameter space in 
$\lambda$ and $g$ whereas $M_{\rm D}=\sqrt 
\xi$ is still of the order of $2 \times 10^{15}$ GeV. Concerning cosmic 
strings contribution to the CMB, we  can see from Fig.~\ref{contribDsugra}
 that it is not constant: it depends strongly on the value of the gauge 
coupling $g$ and the superpotential coupling $\lambda$. For $g\gsim 1$, 
it is not possible that the D-term 
inflationary era lasts $60$ e-foldings. Thus, 
a multiple stage 
inflation is necessary  to solve the horizon problem.  
For $g\gsim 2\times 10^{-2}$,
the cosmic strings contribution is always greater than the WMAP limit, 
thus, it is ruled out.

\begin{figure}[hhh]
\begin{center}
\includegraphics[scale=.6]{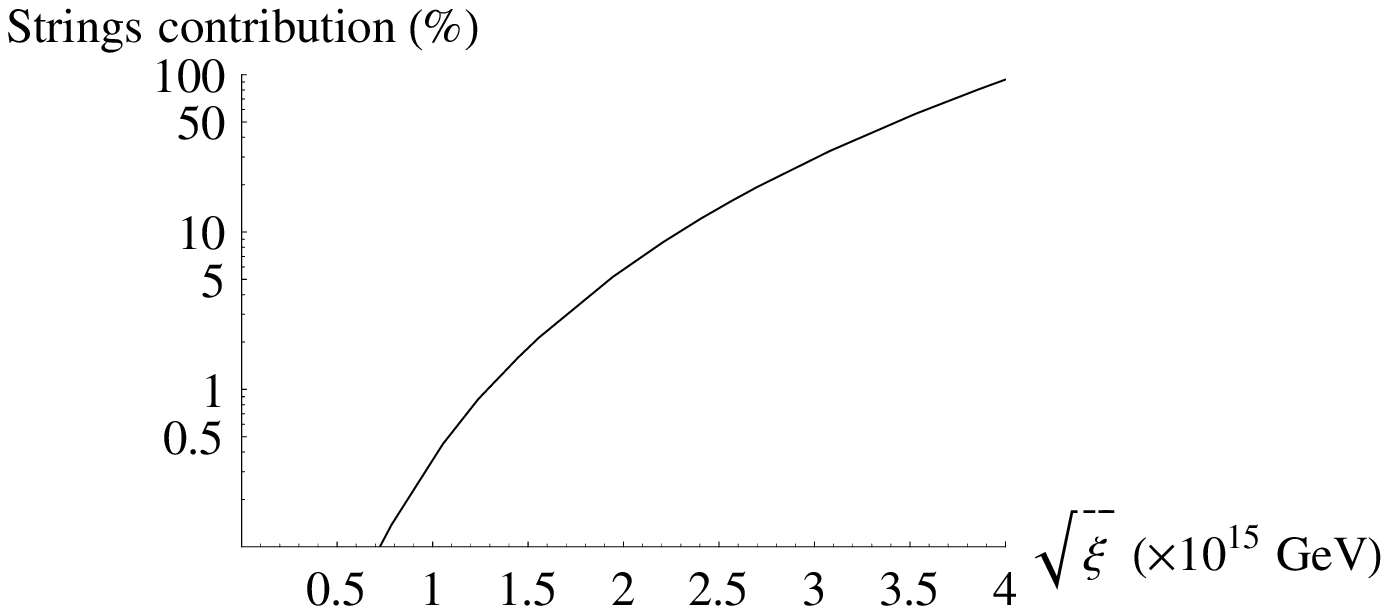}
\includegraphics[scale=.45]{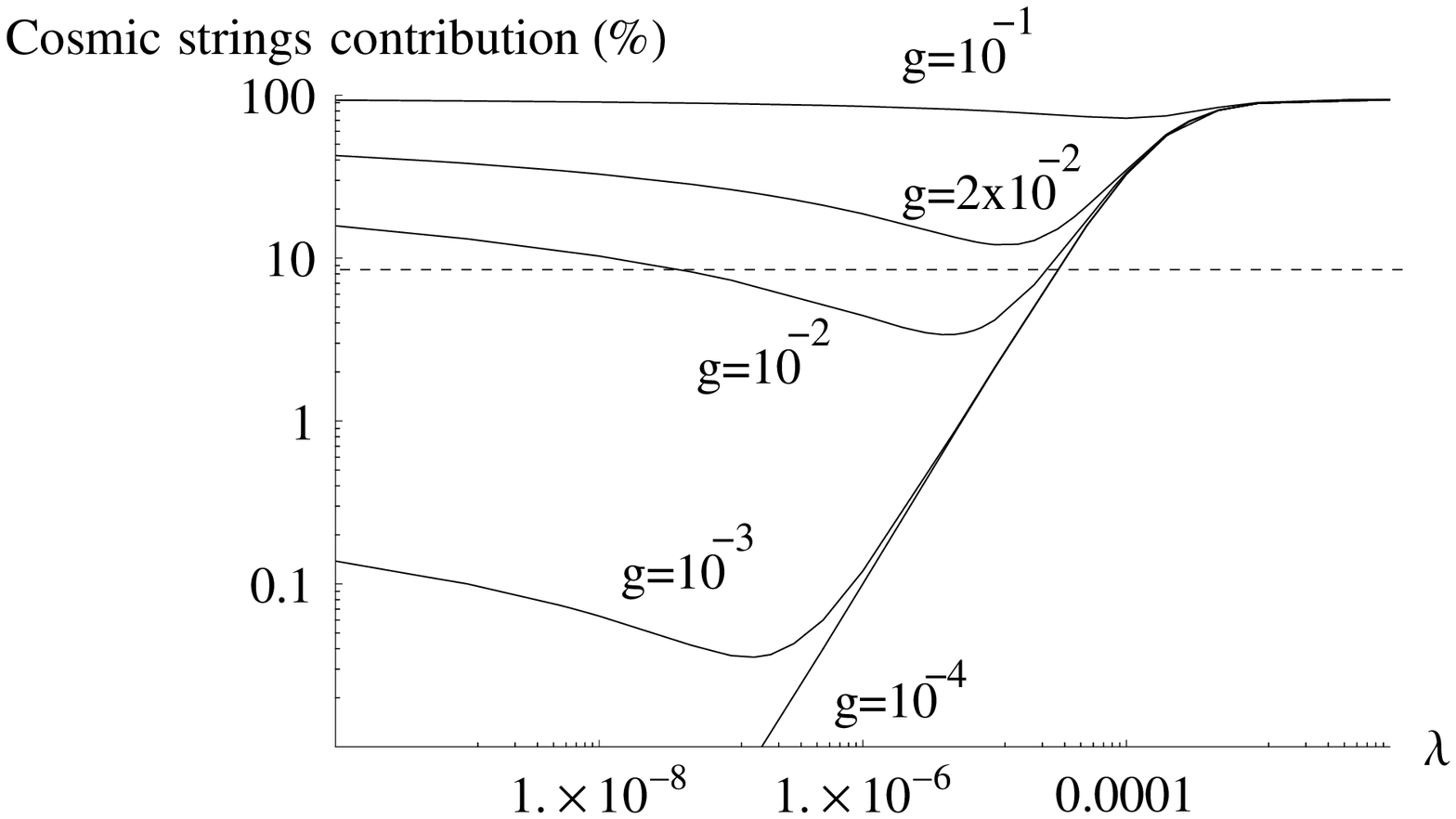}
\caption{On the left, cosmic strings contribution as a function of the 
mass scale $\sqrt\xi$. This holds for all studied values of $g$. On the right, 
cosmic strings contribution to the CMB temperature
anisotropies, as a function of the superpotential coupling  $\lambda$,
for different values of the gauge coupling $g$. The maximal contribution 
alowed by WMAP is represented by a dotted line. The plots are derived in 
the framework of SUGRA.}
\label{contribDsugra}
\end{center}
\end{figure}

For high values of the superpotential coupling $\lambda \gsim 
10^{-3}$, or for small values of the gauge coupling $g$ (namely 
$g\lsim 10^{-4}$), our findings are very similar to the ones 
obtained for D-term inflation within the SUSY context. This is 
expected since in these limits, 
\begin{equation}
{\rm W}\left(\frac{g^2 \xi}{\lambda^2 
M_{\rm Pl}^2}\,z\right)=\frac{|S|^2}{M_{\rm Pl}^2}\ll 1~.
\end{equation}

On the other hand, SUGRA corrections cannot be neglected for small values of
 $\lambda$, in which case the cosmic strings contribution rises again.  It is
 thus possible to specify, for a given value of $g$, the allowed window for
 the superpotential coupling $\lambda$. As we stated earlier, our analysis
 shows that the allowed cosmic strings contribution to the CMB temperature
 anisotropies, as imposed by the WMAP measurements, leads to the constraint
\begin{equation}
g \lsim 2\times 10^{-2}~.
\label{constrg}
\end{equation}
Provided the above condition is satisfied, the CMB constraint sets an
upper bound to the allowed window for the superpotential coupling
 $\lambda$, namely
\begin{equation}
\label{lambda-constr}
\lambda \lsim 3\times 10^{-5}~,
\end{equation}
up to a confidence level of $99\%$. This upper limit on $\lambda$ is the same
as our previous limit found in the framework of SUSY. Note that the upper
bound we found for $\lambda$, Eq.~(\ref{lambda-constr}), is in agreement with
the one found in Ref.\cite{kl} in a slightly different model, where the gauge
coupling $g$ is related to the Yukawa coupling $\lambda$ through the relation
$\lambda={\sqrt 2} g$.  Our constraint on $\lambda$, 
Eq.~(\ref{lambda-constr}), is also in agreement with the finding
$\lambda \lsim {\cal O}(10^{-4}-10^{-5})$ of Ref.~\cite{endo}.

Supergravity corrections induce a lower bound for
$\lambda$ but for $g\lsim 2\times 10^{-3}$ it must be very small and it is
therefore uninteresting, if one wants to avoid fine tuning of the
parameters. However, this lower limit is interesting for $g$ in the range of
$[2\times 10^{-3},~2\times 10^{-2}]$. As an example, for $g=10^{-2}$, the CMB
constraint imposes
\begin{equation}
10^{-8} \lsim\lambda \lsim 3\times 10^{-5}~.
\end{equation}
This allowed window for $\lambda$ shrinks as $g$ goes from
$g=10^{-2}$ to $g=2\times 10^{-2}$.  This is the most important
contribution from supergravity. This fine tuning can be less severe if one 
invokes the curvaton mechanism~\cite{jmprl05}.

As in the case of F-term inflation, it is
also possible to convert this constraint into a constraint on the mass
scale, which is given by the Fayet-Iliopoulos term $\xi$ and we get
\begin{equation}
\sqrt\xi \lsim 2\times 10^{15}~{\rm GeV}~.
\end{equation}
The above constraint on $\xi$ is independent of the value of
the gauge coupling $g$, provided Eq.~(\ref{constrg}) is satisfied.

We would like to bring to the attention of the reader that in the above 
study we have neglected the quantum gravitational effects, i.e., a 
contribution~\cite{lindebook}
\begin{equation}
[\Delta V(|S|)]_{\rm QG}=C_1 {{\rm d}^2 V\over {\rm d}S^2} {V\over 
M_{\rm Pl}^2} + C_2 {V^2\over M_{\rm Pl}^4}
\end{equation} 
(with $C_1, C_2$ numerical coefficients of the order of 1), to the effective 
potential, even though $S_{\rm Q}\sim {\cal O}(10 M_{\rm Pl})$.
Our analysis is however still valid, since the effective potential given in
Eq.~(\ref{VexactDsugra}) satisfies the conditions $V(|S|)\ll M_{\rm Pl}^4$
and $m^2_S={\rm d}^2 V/{\rm d}S^2 \ll M_{\rm Pl}^2$, and thus the quantum
gravitational corrections $[\Delta V(|S|)]_{\rm QG}$ are very small as 
compared to the effective potential 
$V_{\rm eff}^{\rm D-SUGRA}$~\cite{lindebook}.

\section{Conclusions and discussion}

In the context of SUSY GUTs, cosmic strings --- occasionally accompanied by
embedded strings --- are the outcome of SSB schemes, compatible
with high energy physics and cosmology.  As it was explicitely shown in
Ref.~\cite{rjs}, strings are generically formed at the end of a supersymmetric
hybrid inflationary era, which can be either F-term or D-term type, as we
follow the patterns of SSBs from grand unified gauge groups ${\rm G_{GUT}}$
down to the standard model gauge group ${\rm G_{SM}}\times Z_2$.  One should
keep in mind, that there are mechanisms to avoid cosmic strings
formation, which are not considered here since we focus on the most standard
GUT and hybrid inflation models.  The strings forming at the end of inflation
have a mass which is proportional to the inflationary scale. However, current
CMB temperature anisotropies data limit the contribution of cosmic strings to
the angular power spectrum.  The aim of this study is to use the data given
by the realm of cosmology to constrain the parameters of supersymmetric hybrid
inflationary models. 

More precisely, in the framework of generic SUSY GUTs, we study the
consequences of simple hybrid inflation models in the context of some realistic
cosmological scenario. We then compare the observational predictions of these
models in order to constrain the parameter's space. We perform our
calculations within SUSY or SUGRA, whenever this is needed.

For F-term inflation, the symmetry breaking scale $M$ associated with the
inflaton mass scale (and the strings scale) is a slowly varying function of
the superpotential coupling $\kappa$, and of the dimensionality ${\cal N}$ of
the representations to which the scalar components of the chiral Higgs
superfields belong.  For SO(10), under the requirement that R-parity is
conserved down to low energies, $\mathcal{N}=\mathbf{126}$.  For
$\mathrm{E}_6$, the Higgs representations can be $\mathcal{N}=\mathbf{27}$ or
$\mathcal{N}=\mathbf{351}$. The dependence of a measurable quantity (the
cosmic strings contribution) on this discrete parameter ${\cal N}$ can provide
an interesting tool to discriminate among different grand unified gauge groups.

The unique free parameter of the model, the dimensionless coupling $\kappa$,
 should obey two constraints. Namely, it should satisfy the gravitino
 constraint, and it should also obey to the CMB constraint. We found that the
 CMB constraint on the coupling $\kappa$ is the strongest one: it imposes
 $\kappa \lsim 7\times10^{-7} \times (126/\mathcal{N})$, for $\mathcal{N}
 \in \{\mathbf{1}, \mathbf{16},\mathbf{27},\mathbf{126},\mathbf{351}\}$
 whereas the gravitino constraint reads $\kappa \lsim 8\times 10^{-3}$.

The WMAP constraint on the superpotential coupling $\kappa$ can be expressed
into a constraint on the mass scale, namely $M \lsim 2 \times 10^{15}$ GeV.
The value of the inflaton field is of the same order of magnitude, and since
it is below the Planck scale, it implies that global supersymmetry is
sufficient for our analysis. Thus, it is not necessary to take into account
supergravity corrections.

The implications of this finding are quite strong.  Supersymmetric
hybrid inflation was advertised to circumvent the naturalness issue
appearing in the most elegant nonsupersymmetric inflationary
model, i.e.,  chaotic inflation.  Namely,  chaotic inflation 
needed a very small coupling ($\lambda \sim 10^{-14}$). 
Cosmology taught us that supersymmetric hybrid inflation also suffers 
from the fine tuning issue when embedded within the class of cosmological 
models considered here.

In this study we also employed a mechanism where this problem can be
lifted. This is done if we use the curvaton mechanism. In this case another
scalar field, called the curvaton, could generate the initial density
perturbations whereas the inflaton field is only responsible for the dynamics
of the Universe. Within supersymmetric theories such scalar fields are
expected to exist. In this case, the coupling $\kappa$ is only constrained by
the gravitino limit.

We performed the same analysis for D-term inflation. In this case we found
that the value of the inflaton field is of the order of the Planck scale or
higher. One should therefore consider local supersymmetry, taking into account
all the one-loop radiative corrections. To our knowledge, this was never
considered in previous studies.  Our analysis gives a nonconstant contribution
of cosmic strings to the CMB temperature anisotropies, which is strongly
dependent on the gauge coupling $g$ and the superpotential $\lambda$.  We
claim that the D-term inflationary model is still an open possibility, since
it does not always imply that the cosmic strings contribution to the CMB is
above the upper limit allowed by WMAP. To avoid contradiction with the data,
the free parameters of the model are strongly constrained. More precisely, we
found that the gauge coupling must satisfy the constraint $g \lsim 2\times
10^{-2}$, and the superpotential coupling must obey the condition $\lambda
\lsim 3\times 10^{-5}$. This is the same limit as the one found in the SUSY
framework. The supergravity corrections also give a lower limit on the value
of this parameter.  As an example, for $g=10^{-2}$ we found $10^{-8} \lsim
\lambda \lsim 10^{-8}$.  This allowed window for $\lambda$ shrinks as $g$ goes
from $g=10^{-2}$ to $g=2\times 10^{-2}$.  The conditions imposed by the CMB
data on the couplings $\lambda, g$ can be expressed as a single constraint on
the Fayet-Iliopoulos term $\xi$, namely $\sqrt\xi \lsim 2\times 10^{15}~{\rm
GeV}$, which remains valid independently of $g$.

To conclude, we would like to emphasise that cosmic strings of the GUT scale
are in agreement with observationnal data and can play a r\^ole in cosmology.
CMB data do not rule out cosmic strings; they impose strong constraints on
their possible contribution.  One should use these constraints to test high
energy physics such as supersymmetric grand unified theories. This is indeed
the philosophy of our paper.

\ack
It is a pleasure to thank P.\ Brax, N.\ Chatillon,
 G. Esposito-Far\`ese, J.\ Garcia-Bellido, R.\ Jeannerot, A.\ Linde,
J.\ Martin,  P.\ Peter,
and C.\ -M.\  Viallet for discussions and comments.

\section{Appendix}\label{ssb}

We list below the SSB schemes compatible with high energy physics and
cosmology, as given in detail in Ref.~\cite{rjs}.  Let us remind to
the reader that every $\stackrel{n}{\longrightarrow}$ represent an SSB
during which there is formation of topological defects.  Their nature
is given by $n$: $1$ for monopoles, $2$ for topological cosmic
strings, $2'$ for embedded strings, $3$ for domain walls. Please note
also that for e.g. $3_{\rm C}~2_{\rm L}~2_{\rm R}~1_{\rm B-L}$ stands
for the group SU(3)$_{\rm C}\times$ SU(2)$_{\rm L}\times$ SU(2) $_{\rm
R}\times$ U(1)$_{\rm B-L}$.  We refer the reader to Ref.~\cite{rjs}
for more details.

We first give the
SSB schemes of ${\rm SO}(10)$ down to the G$_{\rm SM}\times Z_2$.

\begin{equation}
\label{eq:51}
    {
  {\rm SO}(10) \left\{ \begin{array}{cccccc}

    \stackrel{1}{\longrightarrow} & 5   ~1_{\rm V} &
      \stackrel{1}{\longrightarrow}& 3_{\rm C} ~2_{\rm L} ~1_{\rm Z} ~1_{\rm V}
      &\stackrel{2}{\longrightarrow}& {\rm G}_{\rm SM}~Z_2\\
  \stackrel{1}{\longrightarrow}  &   4_{\rm C} ~2_{\rm L}  ~2_{\rm R}   &
\stackrel{}{\longrightarrow}    & \mbox{\rm Eq. (\ref{eq:ps})}
\\
      \stackrel{1,2}{\longrightarrow}  &   4_{\rm C} ~2_{\rm L}   ~2_{\rm R}     ~Z_2^{\rm C}   &  \stackrel{}{\longrightarrow}    & \mbox{{\rm Eq. (\ref{eq:psZ2})}}

\\
  \stackrel{1,2}{\longrightarrow}  &   4_{\rm C}     ~2_{\rm L}     ~1_{\rm R}     ~Z_2^{\rm C}   &  \stackrel{}{\longrightarrow}    &  \cdots  \\
  \stackrel{1}{\longrightarrow}  &   4_{\rm C}     ~2_{\rm L}     ~1_{\rm R}   &  \stackrel{}{\longrightarrow}  &    \cdots  \\
  \stackrel{1,2}{\longrightarrow}  &   3_{\rm C}     ~2_{\rm L}     ~2_{\rm R}     ~1_{\rm B-L}     ~Z_2^{\rm C}   &  \stackrel{}{\longrightarrow}  &    \cdots  \\
  \stackrel{1}{\longrightarrow}  &   3_{\rm C}     ~2_{\rm L}     ~2_{\rm R}     ~1_{\rm B-L}   &  \stackrel{}{\longrightarrow}    &  \cdots  \\
  \stackrel{1}{\longrightarrow}  &   3_{\rm C}     ~2_{\rm L}     ~1_{\rm R}     ~1_{\rm B-L}   &  \stackrel{2}{\longrightarrow}   &   {\rm G}_{\rm SM}  ~Z_2  \\
  \end{array}
  \right.
    }
\end{equation}

where

  \begin{equation}
\label{eq:ps}
\begin{array}{clllcccc}
 4_{\rm C}     ~2_{\rm L}     ~2_{\rm R}   &
\left\{
\begin{array}{cllllccc}
\stackrel{1}{\longrightarrow} & 3_{\rm C} ~2_{\rm L} ~2_{\rm R} ~1_{\rm
B-L} & \left\{
\begin{array}{cllllccc}
 \stackrel{1}{\longrightarrow}  &   3_{\rm C}     ~2_{\rm L}     ~1_{\rm R}     ~1_{\rm B-L}   &  \stackrel{2}{\longrightarrow}  &   {\rm G}_{\rm SM}  ~Z_2   \\
  \stackrel{2',2}{\longrightarrow}  &   {\rm G}_{\rm SM} ~Z_2\\
 \end{array}
\right.
\\
  \stackrel{1}{\longrightarrow}  &   4_{\rm C}     ~2_{\rm L}     ~1_{\rm R}   &
\left\{
\begin{array}{cllllccc}
  \stackrel{1}{\longrightarrow}  &   3_{\rm C}     ~2_{\rm L}     ~1_{\rm R}     ~1_{\rm B-L}   &   \stackrel{2}{\longrightarrow}  &   {\rm G}_{\rm SM} ~Z_2\\
 \stackrel{2',2}{\longrightarrow}  &   {\rm G}_{\rm SM} ~Z_2\\
 \end{array}
\right.
\\
  \stackrel{1}{\longrightarrow}  &   3_{\rm C}     ~2_{\rm L}     ~1_{\rm R}     ~1_{\rm B-L}   &  ~~~~\stackrel{2}{\longrightarrow}      ~~{\rm G}_{\rm SM} ~Z_2\\
  \end{array}
\right.
\end{array}
\end{equation}
and
\begin{equation}
\label{eq:psZ2}
\begin{array}{clllcccc}
  4_{\rm C}     ~2_{\rm L}     ~2_{\rm R}     ~Z_2^{\rm C}   &
\left\{
\begin{array}{cllllccc}

 \stackrel{1 }{\longrightarrow}  &    3_{\rm C}     ~2_{\rm L}     ~2_{\rm R}     ~1_{\rm B-L}     ~Z_2^{\rm C}   &
\left\{
\begin{array}{cllllccc}
\stackrel{3}{\longrightarrow}  &    3_{\rm C}     ~2_{\rm L}     ~2_{\rm R}     ~1_{\rm B-L}   &   \stackrel{}{\longrightarrow}   &   \cdots  \\
 \stackrel{1,3}{\longrightarrow}  &   3_{\rm C}     ~2_{\rm L}     ~1_{\rm R}     ~1_{\rm B-L}   &  \stackrel{2}{\longrightarrow}  &   {\rm G}_{\rm SM}  ~ Z_2 \\
  \end{array}
\right.
\\
  \stackrel{1}{\longrightarrow}  &   4_{\rm C}     ~2_{\rm L}     ~1_{\rm R}     ~Z_2^{\rm C}   &
\left\{
\begin{array}{cllllccc}
  \stackrel{3}{\longrightarrow}  &   4_{\rm C}     ~2_{\rm L}     ~1_{\rm R}   &  {\longrightarrow}  &   \cdots  \\
  \stackrel{1,3}{\longrightarrow}  &   3_{\rm C}     ~2_{\rm L}     ~1_{\rm R}     ~1_{\rm B-L}   &  \stackrel{2}{\longrightarrow}  &  {\rm G}_{\rm SM} ~Z_2\\
  \end{array}
\right.\\
\stackrel{3}{\longrightarrow}  &   4_{\rm C}     ~2_{\rm L}     ~2_{\rm R}
&{\longrightarrow}   ~~~\mbox{\rm Eq. (\ref{eq:ps})}   \\
  \stackrel{1}{\longrightarrow}  &   4_{\rm C}     ~2_{\rm L}     ~1_{\rm R}   &  {\longrightarrow}     ~~~\cdots  \\
\stackrel{1,3}{\longrightarrow}  &   3_{\rm C}     ~2_{\rm L}     ~2_{\rm R}     ~1_{\rm B-L}   &  \stackrel{}{\longrightarrow}     ~~~\cdots   \\
  \stackrel{1, 3}{\longrightarrow}  &   3_{\rm C}     ~2_{\rm L}     ~1_{\rm R}     ~1_{\rm B-L}    &  \stackrel{2}{\longrightarrow}     ~~~ {\rm G}_{\rm SM} ~Z_2
 \end{array}
\right.
\end{array}
\end{equation}

We then proceed with the list of the SSB schemes of ${\rm  E}_6$
down to the G$_{\rm SM}\times Z_2$. We first give the SSB patterns where
${\rm  E}_6$ is broken via ${\rm SO}(10) \times {\rm U}(1)$}.

\begin{equation}
    {\begin{array}{cllllll}
{\rm E}_6
\stackrel{1}{\rightarrow} 10 ~1_{\rm V'}  &
  \left\{
    \begin{array}{cllllccccc}
 \stackrel{2}{\longrightarrow}  & 10  & \stackrel{}{\longrightarrow} ~~~~~\cdots \\
 \stackrel{1}{\longrightarrow} & 5  ~1_{\rm V}   ~1_{\rm V'}  & \stackrel{}{\longrightarrow} ~~~ \mbox{{\rm Eq.~(\ref{eq:5})}} \\
 \stackrel{1}{\longrightarrow}  &  5_{\rm F}   ~1_{\rm V}   ~1_{\rm V'}
& \stackrel{}{\longrightarrow} ~~~ \mbox{{\rm Eq.~(\ref{eq:5F})}}
\\
 \stackrel{1}{\longrightarrow}  &  5_{\rm E}   ~1_{\rm V}   ~1_{\rm V'}  & \stackrel{2',2}{\longrightarrow}   ~~{\rm G}_{\rm SM}  ~Z_2 \\
 \stackrel{2}{\longrightarrow} & 5  ~1_{\rm V'}  ~ Z_2   & \stackrel{}{\longrightarrow}   ~~~~\cdots \\
 \stackrel{1,2}{\longrightarrow} & 5  ~1_{\rm V}  & \stackrel{}{\longrightarrow}  ~~~~~\cdots \\
  \stackrel{1}{\longrightarrow}  &  5_{\rm F}   ~1_{\rm V} & \stackrel{2',2}{\longrightarrow}  ~~{\rm G}_{\rm SM}  ~ Z_2 \\
\stackrel{1}{\longrightarrow} &  {\rm G}_{\rm SM}   ~1_{\rm V}  & \stackrel{2}{\longrightarrow}   ~~{\rm G}_{\rm SM}  ~Z_2 \\
 \stackrel{1,2}{\longrightarrow} &  {\rm G}_{\rm SM}   ~1_{\rm V'}  ~Z_2  & \stackrel{2}{\longrightarrow}   ~~{\rm G}_{\rm SM}~ Z_2 \\
\stackrel{1,2}{\longrightarrow}& 4_{\rm C} ~2_{\rm L} ~2_{\rm R} ~1_{\rm V'}
& \stackrel{}{\longrightarrow} ~~~ \mbox{{\rm Eq.~(\ref{eq:ps6})}}
\\
\stackrel{1}{\longrightarrow}& 4_{\rm C} ~2_{\rm L} ~2_{\rm R} & \stackrel{}{\longrightarrow} ~~~\mbox{{\rm Eq. (\ref{eq:ps})}}\\
\stackrel{1}{\longrightarrow}& 3_{\rm C} ~2_{\rm L} ~2_{\rm R} ~1_{\rm B-L} ~1_{\rm V'} & \stackrel{}{\longrightarrow} ~~~~\cdots\\
\stackrel{1}{\longrightarrow}& 3_{\rm C} ~2_{\rm L} ~1_{\rm R} ~1_{\rm B-L} ~1_{\rm V'} & \stackrel{}{\longrightarrow} ~~~~\cdots
\\

  \end{array}
  \right.
    \end{array}  }
\end{equation}

with more direct schemes

\begin{equation}
    {
{\rm E}_6\left\{ \begin{array}{clllllcccc}
&\stackrel{1}{\longrightarrow}& 5 ~1_{\rm V} ~1_{\rm V'} &\stackrel{}{\longrightarrow}& \mbox{{\rm Eq.(\ref{eq:5})}}\\
&\stackrel{1}{\longrightarrow}& 5_{\rm F} ~1_{\rm V} ~1_{\rm V'} &\stackrel{}{\longrightarrow}& \mbox{{\rm Eq.(\ref{eq:5F})}}\\
&\stackrel{1}{\longrightarrow}& 5_{\rm E} ~1_{\rm V} ~1_{\rm V'} &\stackrel{2',2}{\longrightarrow}& {\rm G}_{\rm SM}~Z_2\\
&\stackrel{1}{\longrightarrow}& 5 ~1_{\rm V}  &\stackrel{}{\longrightarrow}& \cdots\\
&\stackrel{1}{\longrightarrow}& 5 ~1_{\rm V'}  &\stackrel{}{\longrightarrow}& \cdots\\
&\stackrel{1}{\longrightarrow}& 5_{\rm F} ~1_{\rm V}  &\stackrel{2',2 }{\longrightarrow}& {\rm G}_{\rm SM} ~Z_2\\
&\stackrel{1}{\longrightarrow}& 4_{\rm C} ~2_{\rm L} ~2_{\rm R}
~1_{\rm V'} &\stackrel{}{\longrightarrow}& \mbox{\rm Eq. (\ref{eq:ps6})}\\
&\stackrel{1}{\longrightarrow}& 4_{\rm C} ~2_{\rm L} ~2_{\rm R}
&\stackrel{}{\longrightarrow}& \mbox{{\rm Eq.~(\ref{eq:ps})}}\\
&\stackrel{1}{\longrightarrow}& 4_{\rm C} ~2_{\rm L} ~1_{\rm R}&\stackrel{}{\longrightarrow}& \cdots\\
&\stackrel{1}{\longrightarrow}&
4_{\rm C} ~2_{\rm L} ~1_{\rm R} ~1_{\rm V'}
&\stackrel{}{\longrightarrow}& \cdots\\
&\stackrel{1}{\longrightarrow}& 3_{\rm C} ~2_{\rm L}
~2_{\rm R} ~1_{\rm B-L} ~1_{\rm V'} &\stackrel{}{\longrightarrow}&
\cdots\\
&\stackrel{1}{\longrightarrow}& 3_{\rm C} ~2_{\rm L} ~1_{\rm R} ~1_{\rm
B-L} ~1_{\rm V'} &\stackrel{}{\longrightarrow}& \cdots\\
&\stackrel{1}{\longrightarrow}& 3_{\rm C} ~2_{\rm L} ~1_{\rm R} ~1_{\rm
B-L} &\stackrel{2}{\longrightarrow}& {\rm G}_{\rm SM}~Z_2\\
&\stackrel{1}{\longrightarrow}& {\rm G}_{\rm SM} ~1_{\rm V}  &\stackrel{2}{\longrightarrow}& {\rm G}_{\rm SM}~Z_2 \\
&\stackrel{1,2}{\longrightarrow}& {\rm G}_{\rm SM} ~1_{\rm V'} ~Z_2 &\stackrel{2}{\longrightarrow}& {\rm G}_{\rm SM}~Z_2 \\
  \end{array}
  \right.
    }
\end{equation}

where

\begin{equation}
\label{eq:5}
\begin{array}{clllllccc}
 5  ~1_{\rm V}   ~1_{\rm V'}
& \left\{
\begin{array}{clllllccc}
\stackrel{2}{\longrightarrow} & 5  ~1_{\rm V'}  ~Z_2
& \left\{
\begin{array}{clllllccc}
 \stackrel{1}{\longrightarrow} &  {\rm G}_{\rm SM}   ~1_{\rm V'}  ~ Z_2  & \stackrel{2}{\longrightarrow} &  {\rm G}_{\rm SM}  ~Z_2 \\
\end{array}
  \right.
\\
\stackrel{1}{\longrightarrow} &  {\rm G}_{\rm SM}   ~1_{\rm V}   ~1_{\rm V'}
& \left\{
\begin{array}{cllllllccc}
\stackrel{2}{\longrightarrow} &  {\rm G}_{\rm SM}   ~1_{\rm V}  & \stackrel{2}{\longrightarrow} &  {\rm G}_{\rm SM}  ~ Z_2  \\
 \stackrel{2}{\longrightarrow} &  {\rm G}_{\rm SM}   ~1_{\rm V'}  ~Z_2  & \stackrel{2}{\longrightarrow} &  {\rm G}_{\rm SM}  ~ Z_2 \\
\end{array}
  \right.
\\
 \stackrel{2}{\longrightarrow} & 5  ~1_{\rm V} & \stackrel{}{\longrightarrow}   ~~~~\cdots  \\
\end{array}
  \right.\end{array}
\end{equation}

\begin{equation}
\label{eq:5F}
\begin{array}{clllllccc}
 5_{\rm F}   ~1_{\rm V}   ~1_{\rm V'}
& \left\{
\begin{array}{cllllccc}
~\stackrel{2}{\longrightarrow} &  ~~5_{\rm F}   ~1_{\rm V}   & \stackrel{2',2}{\longrightarrow}   & {\rm G}_{\rm SM} ~Z_2 \\
 ~\stackrel{2',2}{\longrightarrow} &  ~{\rm G}_{\rm SM} ~Z_2 \\
\end{array}
  \right.
\end{array}
\end{equation}

\begin{equation}
\label{eq:ps6}
\begin{array}{clllllccc}
4_{\rm C} ~2_{\rm L} ~2_{\rm R} ~1_{\rm V'}
&\left\{
\begin{array}{cllllccc}
\stackrel{2}{\longrightarrow}& 4_{\rm C} ~2_{\rm L} ~2_{\rm R} & \stackrel{}{\longrightarrow} ~~~\mbox{{\rm Eq.~(\ref{eq:ps})}} \\
\stackrel{1}{\longrightarrow}& 3_{\rm C} ~2_{\rm L} ~1_{\rm R} ~1_{\rm B-L} ~1_{\rm V'}
& \left\{
\begin{array}{cllllllccc}
\stackrel{2}{\longrightarrow}& 3_{\rm C}  ~2_{\rm L} ~1_{\rm R} ~1_{\rm B-L} &\stackrel{2}{\longrightarrow}&  {\rm G}_{\rm SM}~Z_2\\
\stackrel{2',2}{\longrightarrow}& {\rm G}_{\rm SM} 1_{\rm V'}~Z_2 &\stackrel{2}{\longrightarrow}& {\rm G}_{\rm SM}~Z_2\\
\stackrel{2',2}{\longrightarrow}& {\rm G}_{\rm SM}~Z_2\\
  \end{array}
  \right.
\\
\stackrel{1}{\longrightarrow}& 3_{\rm C} ~2_{\rm L} ~2_{\rm R} ~1_{\rm B-L} ~1_{\rm V'}
& \left\{
\begin{array}{cllllccc}
\stackrel{1}{\longrightarrow}& 3_{\rm C} ~2_{\rm L} ~1_{\rm R} ~1_{\rm B-L} ~1_{\rm V'} &\stackrel{}{\longrightarrow}& \cdots\\
\stackrel{2}{\longrightarrow}& 3_{\rm C} ~2_{\rm L} ~2_{\rm R} ~1_{\rm B-L}&\stackrel{}{\longrightarrow}& \cdots\\
\stackrel{2',2}{\longrightarrow}& {\rm G}_{\rm SM} 1_{\rm V'}~Z_2 &\stackrel{2}{\longrightarrow}& {\rm G}_{\rm SM}~Z_2\\
\stackrel{1,2}{\longrightarrow}& 3_{\rm C} ~2_{\rm L} ~1_{\rm R} ~1_{\rm B-L} &\stackrel{2}{\longrightarrow}&  {\rm G}_{\rm SM}~Z_2\\
\stackrel{2',2}{\longrightarrow}& {\rm G}_{\rm SM}~Z_2\\
  \end{array}
  \right.
\\
\stackrel{1}{\longrightarrow}& 4_{\rm C} ~2_{\rm L} ~1_{\rm R} ~1_{\rm V'}
& \left\{
\begin{array}{cllllccc}
 \stackrel{2}{\longrightarrow}& 4_{\rm C} ~2_{\rm L} ~1_{\rm R}  &\stackrel{}{\longrightarrow}& \cdots\\
\stackrel{1}{\longrightarrow}& 3_{\rm C} ~2_{\rm L} ~1_{\rm R} ~1_{\rm B-L} ~1_{\rm V'}  &\stackrel{}{\longrightarrow}& \cdots\\
\stackrel{2',2}{\longrightarrow}& {\rm G}_{\rm SM} 1_{\rm V'}~Z_2 &\stackrel{2}{\longrightarrow}& {\rm G}_{\rm SM}~Z_2\\
\stackrel{1,2}{\longrightarrow}& 3_{\rm C} ~2_{\rm L} ~1_{\rm R} ~1_{\rm B-L} &\stackrel{2}{\longrightarrow}& {\rm G}_{\rm SM}~Z_2 \\
\stackrel{2}{\longrightarrow}& {\rm G}_{\rm SM}~Z_2\\
 \end{array}
  \right.
\\
\stackrel{1,2}{\longrightarrow}& {\rm G}_{\rm SM} 1_{\rm V'}~Z_2 & \stackrel{2}{\longrightarrow} ~~~{\rm G}_{\rm SM}~Z_2\\
\stackrel{1,2}{\longrightarrow}& 3_{\rm C} ~2_{\rm L} ~2_{\rm R} ~1_{\rm B-L} & \stackrel{}{\longrightarrow} ~~~\cdots\\
\stackrel{1}{\longrightarrow}& 3_{\rm C} ~2_{\rm L} ~1_{\rm R} ~1_{\rm B-L} & \stackrel{2}{\longrightarrow} ~~~{\rm G}_{\rm SM}~Z_2\\
  \end{array}
  \right.\end{array}
\end{equation}

We proceed with the allowed SSB patterns of ${\rm E}_6$ down to the
 G$_{\rm SM}\times Z_2$ via ${\rm SU}(3)_{\rm C} \times {\rm
 SU}(3)_{\rm L} \times {\rm SU}(3)_{\rm R}$.

\begin{equation}
{\rm E}_6
\stackrel{0}{\rightarrow} 3_{\rm C} ~3_{\rm L} ~3_{\rm R}
  \left\{
    \begin{array}{cllllll}
\stackrel{1}{\longrightarrow}& 3_{\rm C} ~2_{\rm L} ~2_{\rm (R)} ~1_{\rm Y_{(R)}} ~1_{\rm Y_L} & \stackrel{}{\longrightarrow} & \mbox{\rm Eq.~(\ref{eq:32211})}\\
 \stackrel{1}{\longrightarrow} & 3_{\rm C} ~3_{\rm L} ~2_{\rm (R)} 1_{\rm (Y_R)} & \stackrel{}{\longrightarrow}&  \mbox{\rm Eq.~(\ref{eq:33L21})}
\\
\stackrel{1}{\longrightarrow}& 3_{\rm C} ~2_{\rm L} ~3_{\rm R} ~1_{\rm Y_L} & \stackrel{}{\longrightarrow} & \mbox{\rm Eq.~(\ref{eq:3321})}\\
\stackrel{1}{\longrightarrow}& 3_{\rm C} ~3_{\rm L} ~1_{\rm (R)} ~1_{\rm Y_{(R)}} & \stackrel{}{\longrightarrow} & \mbox{\rm Eq.~(\ref{eq:3311})}\\
\stackrel{1}{\longrightarrow}& 3_{\rm C} ~2_{\rm L} ~1_{\rm (R)} ~1_{\rm Y_{(R)}} ~1_{\rm Y_L}& \stackrel{}{\longrightarrow} & \cdots\\
\stackrel{1}{\longrightarrow}& 3_{\rm C} ~2_{\rm L} ~2_{\rm (R)} ~1_{\rm B-L}
& \stackrel{}{\longrightarrow} & \cdots\\
\stackrel{1}{\longrightarrow}& 3_{\rm C} ~2_{\rm L} ~1_{\rm (R)} ~1_{\rm B-L}
& \stackrel{2}{\longrightarrow}  & {\rm G_{SM}}~Z_2\\
  \end{array}
  \right.
\end{equation}
with  more direct breakings
\begin{equation}
    {
{\rm E}_6\left\{ \begin{array}{clllllcccc} \stackrel{1}{\longrightarrow}& 3_{\rm C} ~2_{\rm L} ~2_{\rm (R)} ~1_{\rm Y_L} ~1_{\rm Y_{(R)}} &\stackrel{}{\longrightarrow}& \mbox{\rm Eq.~(\ref{eq:32211})}\\
\stackrel{1}{\longrightarrow}& 3_{\rm C} ~3_{\rm L} ~2_{\rm (R)} ~1_{\rm Y_{(R)}} &\stackrel{}{\longrightarrow}& \mbox{\rm Eq.~(\ref{eq:33L21})}  \\
\stackrel{1}{\longrightarrow}& 3_{\rm C} ~2_{\rm L} ~3_{\rm R} ~1_{\rm Y_L} &\stackrel{}{\longrightarrow}& \mbox{\rm Eq.~(\ref{eq:3321})} \\
\stackrel{1}{\longrightarrow}& 3_{\rm C} ~2_{\rm L} ~1_{\rm (R)} ~1_{\rm Y_L} ~1_{\rm Y_{(R)}} &\stackrel{}{\longrightarrow}& \cdots\\
\stackrel{1}{\longrightarrow}& 3_{\rm C} ~2_{\rm L} ~2_{\rm (R)} ~1_{\rm B-L} &\stackrel{}{\longrightarrow}& \cdots\\
\stackrel{1}{\longrightarrow}& 3_{\rm C} ~2_{\rm L} ~1_{\rm (R)} ~1_{\rm B-L} &\stackrel{2}{\longrightarrow}& {\rm G}_{\rm SM}~Z_2\\
  \end{array}
  \right.
    }
\end{equation}
where
 \begin{equation}
\label{eq:32211}
    {
3_{\rm C} ~2_{\rm L} ~2_{\rm (R)} ~1_{\rm Y_L} ~1_{\rm Y_{(R)}}
\left\{
\begin{array}{llllll}
\stackrel{1}{\longrightarrow}&  3_{\rm C} ~2_{\rm L} ~2_{\rm (R)} ~1_{\rm B-L}
&\left\{
\begin{array}{lllll}
\stackrel{1}{\longrightarrow}& 3_{\rm C} ~2_{\rm L} ~1_{\rm (R)} ~1_{\rm B-L} &
\stackrel{2}{\longrightarrow} & {\rm G_{SM}}~Z_2\\
 \stackrel{2',2}{\longrightarrow} & {\rm G_{SM}}~Z_2\\
 \end{array}
  \right.
\\
 \stackrel{1}{\longrightarrow}& 3_{\rm C} ~2_{\rm L} ~1_{\rm (R)} ~1_{\rm Y_{(R)}} ~1_{\rm Y_L}
&\left\{
\begin{array}{lllll}
\stackrel{2}{\longrightarrow}& 3_{\rm C} ~2_{\rm L} ~1_{\rm (R)}
~1_{\rm B-L} & \stackrel{2}{\longrightarrow} &  {\rm G_{SM}}~Z_2\\
 \stackrel{2}{\longrightarrow} & {\rm G_{SM}}~Z_2\\
 \end{array}
  \right.
\\
\stackrel{1,2}{\longrightarrow}& 3_{\rm C} ~2_{\rm L} ~1_{\rm (R)} ~1_{\rm B-L}
& \stackrel{2}{\longrightarrow}  ~~~~{\rm G_{SM}}~Z_2\\
\stackrel{2',2}{\longrightarrow}  & {\rm G_{SM}}~Z_2\\
  \end{array}
  \right.
    }
\end{equation}
  \begin{equation}
\label{eq:33L21}
\begin{array}{llllll}
3_{\rm C} ~3_{\rm L} ~2_{\rm (R)} 1_{\rm (Y_R)}
& \left\{
\begin{array}{clllllccccc}
\stackrel{1}{\longrightarrow}& 3_{\rm C} ~2_{\rm L} ~2_{\rm (R)} ~1_{\rm Y_{(R)}} ~1_{\rm Y_L}  &\stackrel{}{\longrightarrow}& \mbox{\rm Eq.~(\ref{eq:32211})} \\
 \stackrel{1}{\longrightarrow}& 3_{\rm C} ~2_{\rm L} ~1_{\rm (R)} ~1_{\rm Y_{(R)}} ~1_{\rm Y_L} &\stackrel{}{\longrightarrow}& \cdots \\
\stackrel{1}{\longrightarrow}& 3_{\rm C} ~2_{\rm L} ~2_{\rm (R)} ~1_{\rm B-L}  &\stackrel{}{\longrightarrow}& \cdots\\
\stackrel{1}{\longrightarrow}& 3_{\rm C} ~2_{\rm L} ~1_{\rm (R)} ~1_{\rm B-L} &\stackrel{2}{\longrightarrow}& {\rm G}_{\rm SM}~Z_2\\
\stackrel{2',2}{\longrightarrow}& {\rm G}_{\rm SM}~Z_2\\
  \end{array}
  \right.
\end{array}
\end{equation}
  \begin{equation}
\label{eq:3321}
\begin{array}{llllll}
3_{\rm C} ~2_{\rm L} ~3_{\rm R} ~1_{\rm Y_L} &\left\{
    \begin{array}{lllll}
\stackrel{1}{\longrightarrow}& 3_{\rm C} ~2_{\rm (R)} ~2_{\rm L} ~1_{\rm Y_L} ~1_{\rm Y_{(R)}} &\stackrel{}{\longrightarrow}&  \mbox{\rm Eq.~(\ref{eq:32211})} \\
\stackrel{2'}{\longrightarrow}& 3_{\rm C} ~2_{\rm (R)}  ~2_{\rm L} ~1_{\rm B-L} &\stackrel{}{\longrightarrow}& \cdots\\
\stackrel{1}{\longrightarrow}& 3_{\rm C} ~2_{\rm L} ~1_{\rm (R)} ~1_{\rm Y_L} ~1_{\rm Y_{(R)}} &\stackrel{}{\longrightarrow}& \cdots\\
\stackrel{1}{\longrightarrow}& 3_{\rm C} ~2_{\rm L} ~1_{\rm (R)} ~1_{\rm B-L} &\stackrel{2}{\longrightarrow}& {\rm G}_{\rm SM}~Z_2\\
\stackrel{2',2}{\longrightarrow}& {\rm G}_{\rm SM}~Z_2\\
  \end{array}
  \right.
\end{array}
\end{equation}
and
 \begin{equation}
\label{eq:3311}
\begin{array}{llllll}
3_{\rm C} ~3_{\rm L} ~1_{\rm (R)} ~1_{\rm Y_{(R)}}&\left\{
\begin{array}{clllll}
\stackrel{1}{\longrightarrow}& 3_{\rm C} ~2_{\rm L} ~1_{\rm Y_L} ~1_{\rm (R)} ~1_{\rm Y_{(R)}} &\stackrel{}{\longrightarrow} & \cdots\\
\stackrel{2'}{\longrightarrow}& 3_{\rm C} ~2_{\rm L} ~1_{\rm (R)} ~1_{\rm B-L} &
\stackrel{2}{\longrightarrow} & {\rm G_{SM}}~Z_2\\
\stackrel{2',2}{\longrightarrow}&  {\rm G_{SM}}~Z_2\\
  \end{array}
  \right.
\end{array}
\end{equation}
Finally, we list below the SSB schemes of E$_6$ down to the G$_{\rm
SM}\times Z_2$ via ${\rm SU}(6) \times {\rm SU}(2)$. These are:
\begin{equation}
\label{eq:62L}
\begin{array}{lll}
\begin{array}{ccc}
{\rm E}_6  &\stackrel{0}{\longrightarrow}  & 6 ~2_{\rm L} \\ \\
&\mbox{or}
&{\rm E}_6
\end{array}
&\left\{
    \begin{array}{cllll}
\stackrel{1}{\longrightarrow}& 3_{\rm C} ~3_{\rm R} ~2_{\rm L} ~1_{\rm Y_L}
& \stackrel{}{\longrightarrow}  & \mbox{\rm Eq. (\ref{eq:3321})} \\
\stackrel{1}{\longrightarrow}& 4_{\rm C} ~2_{\rm L} ~2_{\rm R} ~1_{\rm V'}
&\stackrel{}{\longrightarrow} & \mbox{\rm Eq. (\ref{eq:ps6})}\\
\stackrel{0}{\longrightarrow}& 4_{\rm C} ~2_{\rm L} ~2_{\rm R} &\stackrel{}{\longrightarrow}& \cdots\\
\stackrel{1}{\longrightarrow}& 4_{\rm C} ~2_{\rm L} ~1_{\rm R} ~1_{\rm V'} &\stackrel{}{\longrightarrow}& \cdots\\
\stackrel{1}{\longrightarrow}& 4_{\rm C} ~2_{\rm L}~1_{\rm R} &\stackrel{}{\longrightarrow}& \cdots\\
\stackrel{1}{\longrightarrow}& 3_{\rm C} ~2_{\rm L} ~2_{\rm (R)} ~1_{\rm B-L} &\stackrel{}{\longrightarrow}& \cdots\\
\stackrel{1}{\longrightarrow}& 3_{\rm C} ~2_{\rm L} ~1_{\rm (R)} ~1_{\rm Y_L} ~1_{\rm Y_{(R)}} &\stackrel{}{\longrightarrow}& \cdots\\
\stackrel{1}{\longrightarrow}& 3_{\rm C} ~2_{\rm L} ~1_{\rm (R)} ~1_{\rm B-L} &\stackrel{2}{\longrightarrow}& {\rm G}_{\rm SM}~Z_2\\
\end{array}
  \right.\\
\end{array}
\end{equation}
and
\begin{equation}
\begin{array}{clllll}
\begin{array}{ccc}
{\rm E}_6  &\stackrel{0}{\longrightarrow}  & 6 ~2_{\rm R} \\ \\
&\mbox{or}
&{\rm E}_6
\end{array}
& {
\left\{
    \begin{array}{clllll}
 \stackrel{1}{\longrightarrow}& 4_{\rm C} ~2_{\rm L} ~2_{\rm R} ~1_{V'}
 & \stackrel{}{\longrightarrow} ~~~~\mbox{\rm Eq. (\ref{eq:ps6})} \\
\stackrel{1}{\longrightarrow}& 4_{\rm C} ~2_{\rm L} ~1_{\rm R} ~1_{\rm
 V'} & \stackrel{}{\longrightarrow}~~~~ \cdots\\
 \stackrel{0}{\longrightarrow}& 4_{\rm C} ~2_{\rm L} ~2_{\rm R} &
 \stackrel{}{\longrightarrow} ~~~~ \cdots \\
\stackrel{1}{\longrightarrow}& 4_{\rm C} ~2_{\rm L} ~1_{\rm R} &
 \stackrel{}{\longrightarrow}~~~~ \cdots\\
  \stackrel{1}{\longrightarrow}& 3_{\rm C} ~2_{\rm L} ~2_{\rm R} ~1_{\rm
 B-L} ~1_{\rm V'} & \stackrel{}{\longrightarrow}~~~~ \cdots\\
 \stackrel{1}{\longrightarrow}& 3_{\rm C} ~2_{\rm L} ~2_{\rm R} ~1_{\rm
 B-L} & \stackrel{}{\longrightarrow}~~~~ \cdots\\
 \stackrel{1}{\longrightarrow}& 3_{\rm C} ~2_{\rm L} ~1_{\rm R} ~1_{\rm
 B-L} & \stackrel{2}{\longrightarrow}~~~~ {\rm G}_{\rm SM}~Z_2
\end{array}
  \right.
 }\\  \end{array}
\end{equation}

As it was shown in Ref.~\cite{rjs}, the SSB schemes of SU(6) and SU(7)
down to the standard model which could accommodate an inflationary era
with no defect (of any kind) at later times are inconsistent with
proton lifetime measurements and minimal SU(6) and SU(7) does not
predict neutrino masses.  Thus, these models are incompatible with
high energy physics phenomenology.


\section*{References}
\begin{thebibliography}{99}
\bibitem{SK} 
Y.~Fukuda, {\it et al.}  [Super-Kamiokande Collaboration], Phys.\ Rev.\
Lett.\ {\bf 81}, 1562 (1998).

\bibitem{SNO} Q.~R.~Ahmad, {\it et al.}  [SNO Collaboration], Phys.\ Rev.\
Lett.\ {\bf 87}, 071301 (2001).

\bibitem{kamland} 
K. Eguchi, {\it et al.} [KamLAND Collaboration], Phys.\ Rev.\ Lett.\
{\bf 90}, 021802 (2003).

\bibitem{kibble} 
T.\ W.\ B.\ Kibble, J.\ Phys.\ A{\bf 9}, 387 (1976).

\bibitem{rjs} R.\ Jeannerot, J.\ Rocher, and M. \ Sakellariadou,
Phys.\ Rev. D{\bf 68}, 103514 (2003).

\bibitem{RJ1}
R.\ Jeannerot, Phys.\ Rev. D{\bf 56}, 6205 (1997).

\bibitem{sugra}
E.\ J.\ Copeland, A.\ R.\ Liddle, D.\ H.\ Lyth, E.\ D.\ Stewart, and D.\ Wands,
Phys.\ Rev. D{\bf 49}, 6410 (1994).

\bibitem{anomalousU1} M.\ Dine, N.\ Seiberg, and E.\ Witten, Nucl.\
Phys.\ B{\bf 289}, 585 (1987).

\bibitem{jm1}
J.~Martin, A.~Riazuelo, and M.~Sakellariadou,
Phys.~Rev.~D {\bf 61}, 083518 (2000).

\bibitem{jm2}
A.\ Gangui, J.~Martin, and M.~Sakellariadou,
Phys.~Rev.~D {\bf 66}, 083502 (2002).

\bibitem{lw2002}
D.\ H.\ Lyth and D.\ Wands, Phys.\ Lett.\ B{\bf524}, 5 (2002).

\bibitem{mt2001}
T.\ Moroi, and T.\ Takahashi, Phys.\ Lett.\ B{\bf522}, 215 (2001), 
Erratum-ibid.\ B{\bf 539}, 303 (2002).

\bibitem{ekm2003}
K.\ Enqvist, S.\ Kasuya, and A.\ Mazumdar, Phys.\ Rev.\ Lett.\ {\bf90}, 091302
(2003).

\bibitem{dl2004} 
K.\ Dimopoulos and D.\ Lyth, Phys.\ Rev.\ D{\bf 69}, 123509 (2004).

\bibitem{mt2002}
T.\ Moroi, and T.\ Takahashi, Phys.\ Rev. D{\bf 66}, 063501 (2002).

\bibitem{shifted} R.\ Jeannerot, S.\ Khalil, G.\ Lazarides, and Q.\
Shafi, JHEP\textbf{0010},012 (2000).

\bibitem{smooth}
G.\ Lazarides, and C.\ Panagiotakopoulos, Phys.\ Rev.\ D{\bf 52}, 559 (1995).

\bibitem{watari}
T.\ Watari, and T.\ Yanagida, Phys.\ Lett.\ B {\bf 589}, 71 (2004).

\bibitem{davis}
J.\ Urrestilla, A.\ Ach\'ucarro, and A.\ C.\ Davis,  
Phys.\ Rev.\ Lett. {\bf 92}, 251302 (2004).

\bibitem{3scale} D.~Austin, E.\ J.~Copeland, and T.\ W.\ B.~Kibble,
Phys.\ Rev.\ D{\bf 48}, 5594 (1993).

\bibitem{jm4} 
J.\ Rocher, and M.\ Sakellariadou, \emph{The nature of cosmic strings
within SUSY GUTs}  (in preparation).

\bibitem{davis2}
A.\ -C.\ Davis, {\sl private communication} (2004).

\bibitem{davis3}
S.\ C.\ Davis, A.\ -C.\ Davis, M.\ Trodden, Phys.\ Lett.\ B {\bf 405}, 257
(1997).

\bibitem{sv}
 M.\ Sakellariadou, and A.\ Vilenkin,  Phys.\ Rev. D{\bf 37}, 885 (1988).

\bibitem{s}
M.\ Sakellariadou, Nucl.\ Phys.\ B{\bf 468}, 319 (1996).

\bibitem{ls2003}
M.\ Landriau, and E.\ P.\ S.\ Shellard, Phys.\ Rev.\ D{\bf 69}, 023003 (2004). 

\bibitem{cw} C.\ Coleman, and E.\ Weinberg, Phys.\ Rev.\ D{\bf 7},
1888 (1973).

\bibitem{DvaShaScha} 
G.\ Dvali, Q.\ Shafi, and R.\ Schaefer, Phys.\ Rev.\ Lett.\ {\bf73},
1886 (1994).

\bibitem{lr}
D.\ Lyth and A.\ Riotto, Phys.\ Rep.\ {\bf 314}, 1 (1999).

\bibitem{dm}
A.\ -C.\ Davis and M.\ Majumdar, Phys.\ Lett.\ B {\bf 460}, 257
(1999).

\bibitem{Lazarides} 
G.\ Lazarides, \emph{Inflationary cosmology}, [arXiv:hep-ph/0111328].

\bibitem{SenoSha}
V.\ N.\ Senoguz, and Q.\ Shafi,  Phys.\ Lett.\ B {\bf 567}, 79 (2003).

\bibitem{cobe}
C.\ L.\ Bennett, et.\ al.\, Astrophys.\ J.\ {\bf 464}, 1 (1996).

\bibitem{pati}
J.\ C.\ Pati, Int.\ J.\ Mod.\ Phys.\ A{\bf 18}, 4135 (2003).

\bibitem{laz}
G.\ Lazarides, R.\ K.\ Schafer, and Q.\ Shafi, Phys.\ Rev.\ D{\bf
56}, 1324 (1997).

\bibitem{gr}
M.\ Yu\ Khlopov, A.\ Linde, Phys.\ Lett.\ B {\bf 138}, 265
(1984).

\bibitem{ss2004}
V.\ N.\ Senoguz, Q.\ Shafi,  Phys.\ Lett.\ B {\bf 82}, 6 (2004).

\bibitem{boom}
 C.\ B.\ Netterfield, et.\ al., Astrophys.\ J.\ {\bf 571}, 604 (2002);
 P.~de~Bernardis, et.\ al., Astrophys.\ J.\ {\bf 564}, 559 (2002).

\bibitem{maxi}
 A.\ T.\ Lee, Astrophys.\ J.\ {\bf 561}, L1 (2001); R.\
Stompor, Astrophys.\ J.\ {\bf 561}, L7 (2001).

\bibitem{dasi}
N.\ W.\ Halverson, et.\ al., Astrophys.\ J.\  {\bf 568}, 38 (2002); 
C.\ Pryke, et.\ al.\ , Astrophys.\ J.\ {\bf  568}, 46 (2002).

\bibitem{bouchet}
F.\ R.\ Bouchet, P.\ Peter, A.\ Riazuelo, and M.\ Sakellariadou,
Phys.\ Rev. D{\bf 65}, 021301 (2002).

\bibitem{wmap}
C.\ L.\ Bennett et al., Astroph.\ J.\ Suppl.\ {\bf 148}, 1 (2003).

\bibitem{pogosian} 
L.\ Pogosian, M.\ Wyman, and I.\ Wasserman, JCAP {\bf 0409}, 008 (2004).

\bibitem{kl}
R.\ Kallosh and A.\ Linde, JCAP {\bf 0310}, 008 (2003).

\bibitem{nilles}
P.\ Nilles, Phys.\ Rep.\ {\bf 110}, 1 (1984).

\bibitem{endo}
M.\ Endo, M.\ Kawasaki, and T.\ Moroi, Phys.\ Lett.\ B\textbf{569}, 73 (2003).

\bibitem{DSUGRA}
P.\ Bin\'etruy and G.\ Dvali, Phys.\ Lett.\ B\textbf{388}, 241 (1996).

\bibitem{halyo}
E.\ Halyo, Phys.\ Lett.\  B\textbf{387}, 43 (1996).

\bibitem{jmprl05}
J.\ Rocher, and M. \ Sakellariadou,  Phys.\ Rev.\ Lett.\ {\bf 94}, 011303 
(2005).

\bibitem{lindebook}
A.\ Linde, {\sl Particle Physics and Inflationary Cosmology}, Contemporary 
Concepts in Physics, {\bf 5} (1990), Hardwood Academic Publishers.

\end{thebibliography}
\end{document}